\documentclass[twocolumn]{aastex701}
\usepackage{xcolor}
\usepackage{amsmath}
\usepackage{lipsum}
\usepackage{comment}
\usepackage{makecell}
\usepackage{hyperref}
\usepackage{csvsimple}
\usepackage{xspace}
\usepackage{longtable,array,booktabs}
\usepackage{subcaption} 
\usepackage{multirow}

\newcommand{\msun}{\,\mathrm{M}_\odot}

\defcitealias{2026ApJ...996...69M}{M26}
\newcommand{\miller}{\citetalias{2026ApJ...996...69M}\xspace}

\begin{document}

\title{A Spectroscopic Census of Cluster White Dwarfs from DESI, LAMOST, and SDSS}

\correspondingauthor{David R. Miller}
\email{drmiller@phas.ubc.ca}

\author[0000-0002-4591-1903]{David R. Miller}
\affiliation{Department of Physics and Astronomy, University of British Columbia, Vancouver, BC V6T 1Z1, Canada}
\email{drmiller@phas.ubc.ca}

\author[0000-0001-9739-367X]{Jeremy Heyl}
\affiliation{Department of Physics and Astronomy, University of British Columbia, Vancouver, BC V6T 1Z1, Canada}
\email{heyl@phas.ubc.ca}

\author[orcid=0009-0005-8565-0858]{Huahui Yan}
\affiliation{School of Physics and Astronomy, Sun Yat-sen University, DaXue Road, Zhuhai 519082, GuangDong, People’s Republic of China.}
\affiliation{CSST Science Center for the Guangdong-Hongkong-Macau Greater Bay Area, Sun Yat-sen University, DaXue Road, Zhuhai 519082, GuangDong, People’s Republic of China}
\email{yanhh8@mail2.sysu.edu.cn}  

\author[orcid=0000-0002-3084-5157]{Chengyuan Li}
\affiliation{School of Physics and Astronomy, Sun Yat-sen University, DaXue Road, Zhuhai 519082, GuangDong, People’s Republic of China.}
\affiliation{CSST Science Center for the Guangdong-Hongkong-Macau Greater Bay Area, Sun Yat-sen University, DaXue Road, Zhuhai 519082, GuangDong, People’s Republic of China}
\email{lichengy5@mail.sysu.edu.cn}

\author[0000-0001-9873-0121]{Pier-Emmanuel Tremblay}
\affiliation{Department of Physics, University of Warwick, Coventry, CV4 7AL, UK}
\email{P.Tremblay@warwick.ac.uk}


\begin{abstract}

Gaia has substantially expanded the sample of candidate white dwarfs (WDs) in open clusters, but many remain without spectroscopic characterization. We leverage the increasingly extensive archives of large public spectroscopic surveys to identify 62 candidate cluster WDs with available spectra from DESI, LAMOST, and SDSS. After discarding sources with very low S/N and unreliable host clusters, 44 candidates remain, including 37 non-magnetic DA WDs, of which 15 had not previously been characterized in the literature for cluster studies. We find that nine are consistent with single-star origins in their proposed host clusters, increasing the combined cluster DA initial--final mass relation (IFMR) sample to 132 WDs. All lie within $2\sigma$ of the best-fit IFMR from the most comprehensive recent cluster WD sample, suggesting these additions do not require recalibrating the empirical IFMR. One is a likely Hyades tidal-tail WD, providing further evidence that dynamical dispersal helps explain the long-standing WD deficit in the nearest cluster. We further compare WDs with spectra from multiple surveys. Among seven Praesepe WDs observed by both DESI and SDSS, DESI gives higher WD masses in six cases, with a mean difference of $0.067\msun$, while the inferred progenitor masses differ by as much as $0.459\msun$. More broadly, nine of ten single-star cluster DAs with DESI spectra lie above the empirical IFMR, with a mean final-mass residual of $0.06\pm0.04\msun$. These results show that survey-dependent differences can strongly affect cluster WD parameters, and should be considered when combining heterogeneous spectroscopy in empirical IFMR studies.

\end{abstract}



\section{Introduction} 
\label{sec:intro}

White dwarfs (WDs) in star clusters provide important constraints on stellar and WD evolution because cluster membership provides independent constraints on their ages and progenitor metallicities, quantities that are generally unavailable for field WDs. Cluster WDs are useful for studying atmospheric evolution \citep{2005ApJ...618L.129K,2006ApJ...643L.127W}, the origin of magnetic WDs \citep{2020ApJ...901L..14C}, binary evolution \citep{2014ApJ...783L...8G,2024ApJ...976..102G,2026ApJ...999..167Y}, and dynamical loss and spatial distribution within clusters \citep{1992AJ....104.1876W,2003ApJ...595L..53F,2007MNRAS.382..915H,2022ApJ...926..132H}. They are also central to constraining the empirical initial--final mass relation (IFMR; \citealt{2000A&A...363..647W,2005MNRAS.361.1131F,2018ApJ...866...21C,2026ApJ...996...69M}), which probes cumulative stellar mass loss and core growth and is a critical input to stellar population synthesis models.

The number of candidate WDs to which these constraints may be applied has grown substantially with the Gaia mission \citep{2016A&A...595A...1G,2018A&A...616A...1G,2021A&A...649A...1G,2023A&A...674A...1G}, which has greatly expanded the known WD population \citep{2021MNRAS.508.3877G} and improved the census of Milky Way star clusters \citep{2020A&A...640A...1C,2021MNRAS.504..356D,2023A&A...673A.114H}. Together, these advances have enabled systematic searches for previously unrecognized cluster WDs \citep{2021ApJ...912..165R,2021A&A...645A..13P,2026ApJ...996...69M}. However, candidate samples have grown faster than spectroscopic characterization, and dedicated follow-up remains valuable but is limited by available telescope time. 

Spectroscopy provides a direct way to characterize these WDs through spectral classification and the identification of magnetic or composite systems. Atmospheric fits to the spectra yield measurements of $T_{\rm eff}$ and $\log g$ independent of photometric estimates \citep{1992ApJ...394..228B,2011ApJ...730..128T}, which can then be mapped to WD masses and cooling ages using evolutionary sequences \citep{2020ApJ...901...93B}. The cluster environment allows these measurements to be combined with cluster properties to test membership and constrain the progenitor lifetime and initial mass. 

Large public spectroscopic surveys provide an efficient way to characterize these growing samples. Over the past two decades, wide-field surveys have produced spectra for tens of thousands of WDs, including many objects without previous spectroscopic classifications \citep{2004ApJ...607..426K,2006ApJS..167...40E,2013ApJS..204....5K,2015MNRAS.454.2787G,2015MNRAS.452..765G,2015MNRAS.446.4078K,2016MNRAS.455.3413K,2019MNRAS.486.2169K,2021MNRAS.507.4646K,2022MNRAS.509.2674G,2023ApJS..268...28T,2024MNRAS.535..254M,2026ApJ..1000..216K,2026ApJ..1006..248K,2026MNRAS.546ag246M,2026MNRAS.tmp.1582S}. These include the Dark Energy Spectroscopic Instrument (DESI; \citealt{2016arXiv161100036D}), the Large Sky Area Multi-Object Fiber Spectroscopic Telescope (LAMOST; \citealt{2012RAA....12.1197C}), and the Sloan Digital Sky Survey (SDSS; \citealt{2000AJ....120.1579Y}). These archives allow candidate cluster WDs to be identified and characterized without new observations, while repeat observations across surveys provide a direct test of the consistency of the inferred WD parameters.

In this work, we use archival spectroscopy from DESI, LAMOST, and SDSS to assemble a census of candidate cluster WDs with available spectra. We first identify and characterize candidate cluster WDs not included in recent spectroscopic studies, using their WD and host properties to assess whether they are consistent with single-star cluster membership. We then examine the retained DA WDs in the context of the empirical IFMR, while magnetic and non-DA objects identified in the census are considered separately in \citet{Miller2026Atypical}. Finally, we use WDs observed by multiple surveys to test the consistency of parameters derived from independent spectra and assess how survey-dependent differences affect cluster WD studies.

We describe the spectroscopic census in Section~\ref{sec:sample}, outline the WD and cluster analysis in Section~\ref{sec:analysis}, present the newly characterized cluster WDs and cross-survey comparison in Section~\ref{sec:results}, and summarize our findings in Section~\ref{sec:conclusions}.


\section{Spectroscopic Census}
\label{sec:sample}

We assemble our candidate cluster WD sample from the open cluster and moving group memberships of \citet{2024A&A...686A..42H} and the Unified Cluster Catalog (UCC; \citealt{2023MNRAS.526.4107P}), excluding globular clusters. We retain moving groups because some have sufficiently well-constrained populations and ages for cluster WD analysis. We do not cut on the reported membership probabilities. WDs can lie well beyond their cluster centers \citep{2022ApJ...926..132H,2022ApJ...926L..24M,2023ApJ...956L..41M}, potentially aided by small natal velocity kicks \citep{2003ApJ...595L..53F,2007MNRAS.382..915H,2009ApJ...695L..20F,2018MNRAS.480.4884E,2026PASP..138i4208F,2026arXiv260920940M,2026arXiv260920941P}. Even well-established members can receive relatively low probabilities from clustering algorithms, so we retain all candidates identified by either catalog and use WD and cluster properties to further assess membership. However, because our search begins from the published membership lists, our census remains incomplete for WDs omitted by the underlying clustering algorithms, including objects displaced sufficiently far from their birth clusters to escape identification as candidate members. Within these membership lists, we select WDs from the high-probability WD sample of \citet{2021MNRAS.508.3877G}, corresponding to $P_{\rm WD}>0.75$. 

We crossmatch this WD candidate list against three large public spectroscopic surveys containing extensive samples of Milky Way stars: DESI DR1 \citep{2026AJ....171..285D}, LAMOST DR11 \citep{2012RAA....12.1197C}, and SDSS DR20 \citep{2026arXiv260726149S}. For WDs with multiple observations within a given survey, we retain the highest S/N spectrum; for sources observed by multiple surveys, we retain one spectrum from each survey. To avoid ambiguity with designations used in previous studies, we assign each WD a standardized name of the form Cluster-WDJHHMMSS$\pm$DDMMSS, based on its adopted host and Gaia DR3 coordinates. Two WDs previously established in the literature as Praesepe members and analyzed as such in \citet{2026ApJ...996...69M} (hereafter \miller), Praesepe-WDJ085947+422112 and Praesepe-WDJ092713+475659, are identified by \citet{2024A&A...686A..42H} as candidate members of HSC~1224. We retain Praesepe as their adopted host here.

The resulting census, presented in Table~\ref{tab:wd_sample}, contains 62 candidate cluster WDs, all with $P_{\rm WD}>0.9$. Praesepe and the Hyades contribute 16 and nine WDs, respectively, together accounting for 40\% of the census and reflecting their proximity and rich WD populations. We visually inspected each available spectrum for suitability for classification and fitting. Four candidates have S/N too low for reliable classification, while HSC1553-WDJ060543+154221 can be classified as a DA but has insufficient S/N for detailed atmospheric analysis. We exclude these five sources from detailed analysis.

\begin{table*}[h]
\tiny
\caption{Candidate cluster white dwarfs with archival spectroscopy.}
\label{tab:wd_sample}
\centering
\begin{tabular}{ccccccccc}
\toprule
Name &
Gaia DR3 source ID &
RA (deg) &
Dec (deg) &
Survey &
Catalog &
Type &
Host & 
Analysis
\\
\midrule
Alessi13-WDJ033301-384025   & 4853382867764646912 & 53.256  & -38.674 & S    & Both & DAH    & Usable   & \citet{Miller2026Atypical} \\
Alessi62-WDJ185530+214138   & 4519349757798439936 & 283.875 & 21.694 & S     & Both & DA     & Usable   & this work \\
Alessi94-WDJ001937+461007   & 391939027303287040  & 4.906   & 46.169 & S     & Both & DA     & Usable   & this work \\
CWNU1095-WDJ101309+524349   & 851411295734572416  & 153.291 & 52.730 & D     & H\&R & DA     & Usable   & this work \\
CWNU1149-WDJ171339+190813   & 4548600336193327872 & 258.417 & 19.137 & D     & UCC  & DA     & Excluded & \ldots \\
CWNU1182-WDJ143159+000216   & 3653404636310252416 & 217.999 & 0.038  & D     & UCC  & DA     & Excluded & \ldots \\
FSR1017-WDJ052436-053512    & 3209308900555979776 & 81.151  & -5.587 & S     & UCC  & DA     & Excluded & \ldots \\
HSC190-WDJ161149+041937     & 4437271867601913472 & 242.954 & 4.327  & D     & Both & DA     & Excluded & \ldots \\
HSC242-WDJ163252+093101     & 4446494085800777984 & 248.220 & 9.517  & D     & H\&R & \ldots & Usable   & \ldots\\
HSC381-WDJ175813+172113     & 4502736137087916032 & 269.554 & 17.354 & S     & H\&R & DA     & Usable   & this work \\
HSC448-WDJ214848-035539     & 2673038386327748224 & 327.202 & -3.928 & D     & H\&R & DA     & Usable   & this work \\
HSC448-WDJ222318+000556     & 2678298278156118144 & 335.827 & 0.099  & D+L   & H\&R & DA     & Usable   & this work \\
HSC453-WDJ164331+045644     & 4434749725365419520 & 250.881 & 4.946  & D     & H\&R & DC     & Excluded & \ldots \\
HSC913-WDJ232411+653053     & 2210180554096762624 & 351.047 & 65.515 & S     & UCC  & DA     & Excluded & \ldots \\
HSC1077-WDJ110036+665948    & 1059327432718366720 & 165.153 & 66.997 & D+S   & UCC  & DA     & Excluded & \ldots\\
HSC1224-WDJ092525+525651    & 1019891219101344256 & 141.356 & 52.948 & D+S   & H\&R & DC     & Usable   & \citet{Miller2026Atypical} \\
HSC1279-WDJ050801+451655    & 205722198972554752  & 77.004  & 45.282 & S     & UCC  & DA     & Excluded & \ldots \\
HSC1340-WDJ041034+160321    & 45297409459299328   & 62.643  & 16.056 & S     & UCC  & DA     & Usable   & this work \\
HSC1542-WDJ060517+145552    & 3345502137417354496 & 91.323  & 14.931 & L     & H\&R & DA     & Excluded & \ldots \\
HSC1542-WDJ061502+044802    & 3317802244056844288 & 93.761  & 4.801  & S     & H\&R & DB     & Excluded & \ldots \\
HSC1553-WDJ060543+154221    & 3345986781525842560 & 91.433  & 15.706 & L     & UCC  & DA$^\dagger$ & Usable & \ldots\\
HSC1555-WDJ084658+241239    & 690326898030489984  & 131.742 & 24.211 & D+S   & UCC  & DC     & Excluded & \ldots \\
HSC1555-WDJ084802+204711    & 661798660099237760  & 132.010 & 20.787 & S     & H\&R & \ldots & Excluded & \ldots \\
HSC1555-WDJ085028+271955    & 692795546450243456  & 132.618 & 27.332 & L+S   & UCC  & DA     & Excluded & \ldots \\
HSC1555-WDJ090403+182751    & 612137052768974848  & 136.015 & 18.464 & S     & H\&R & \ldots & Excluded & \ldots \\
HSC1630-WDJ042914+093722    & 3299641271199703680 & 67.310  & 9.623  & L     & H\&R & DA     & Usable   & this work\\
Hyades-WDJ035200+340726     & 218783542413339648  & 58.001  & 34.124 & L     & H\&R & DA     & Usable   & \miller\\
Hyades-WDJ040342+145928     & 39305036729495936   & 60.926  & 14.991 & L     & Both & DA     & Usable   & \miller \\
Hyades-WDJ040842+292130     & 164297449857329408  & 62.176  & 29.358 & L     & H\&R & DA     & Usable   & this work \\
Hyades-WDJ040929+170754     & 45980377978968064   & 62.371  & 17.132 & L     & UCC  & DA     & Usable   & \miller \\
Hyades-WDJ042355+162114     & 3313606340183243136 & 65.983  & 16.354 & L     & UCC  & DA     & Usable   & \miller \\
Hyades-WDJ042839+165811     & 3313714023603261568 & 67.165  & 16.970 & L     & UCC  & DA     & Usable   & \miller \\
Hyades-WDJ043345+124240     & 3306722607119077120 & 68.438  & 12.711 & L     & UCC  & DA     & Usable   & \miller \\
Hyades-WDJ044023+135845     & 3308403897837092992 & 70.100  & 13.979 & L     & Both & DBA    & Usable   & \citet{Miller2026Atypical} \\
Hyades-WDJ072228+221210     & 866166024521723776  & 110.619 & 22.203 & L+S   & H\&R & DA     & Usable   & this work \\
Melotte111-WDJ121100+225220 & 4001560041148002432 & 182.751 & 22.872 & L     & UCC  & DA     & Usable   & this work \\
Melotte111-WDJ121856+254557 & 4008511467191955840 & 184.734 & 25.766 & L     & Both & DA     & Usable   & \miller \\
M67-WDJ085028+105417        & 598655567036876928  & 132.618 & 10.905 & S     & H\&R & DA     & Usable   & this work \\
M67-WDJ085119+114840        & 604917698073661952  & 132.833 & 11.811 & S     & Both & DA     & Usable   & this work \\
M67-WDJ085134+114217        & 604898490980773888  & 132.895 & 11.705 & D+S   & Both & DB     & Usable   & \citet{Miller2026Atypical} \\
M67-WDJ085425+113829        & 604721293514179712  & 133.605 & 11.642 & D+S   & H\&R & DA     & Usable   & this work \\
NGC3532-WDJ110315-582144    & 5338718261060841472 & 165.813 & -58.362& S     & Both & DA     & Usable   & \miller \\
Pleiades-WDJ035211+245548   & 66697547870378368   & 58.047  & 24.930 & L     & Both & DA     & Usable   & \miller\\
Praesepe-WDJ081656+204945   & 675767959626673152  & 124.234 & 20.829 & L+S   & H\&R & DB     & Usable   & \citet{Miller2026Atypical} \\
Praesepe-WDJ082547+174818   & 662152977721471488  & 126.448 & 17.805 & D+L+S & H\&R & DC     & Usable   & \citet{Miller2026Atypical} \\
Praesepe-WDJ082839+194343   & 662998983199228032  & 127.166 & 19.729 & D+S   & Both & DA     & Usable   & this work \\
Praesepe-WDJ083609+193818   & 662798086105290112  & 129.042 & 19.638 & S     & Both & DA     & Usable   & \miller \\
Praesepe-WDJ083822+175744   & 659335994570713600  & 129.594 & 17.962 & S     & H\&R & \ldots & Usable   & \ldots \\
Praesepe-WDJ083936+193043   & 661270898815358720  & 129.902 & 19.512 & S     & Both & DA     & Usable   & \miller \\
Praesepe-WDJ083947+194611   & 661311267210542080  & 129.947 & 19.770 & D+S   & Both & DA     & Usable   & \miller \\
Praesepe-WDJ083945+200015   & 664325543977630464  & 129.940 & 20.004 & D+S   & Both & DAH    & Usable   & \citet{Miller2026Atypical} \\
Praesepe-WDJ084013+184326   & 659494049367276544  & 130.055 & 18.724 & D+L+S & Both & DA     & Usable   & \miller \\
Praesepe-WDJ084028+194334   & 661297901272035456  & 130.117 & 19.726 & L+S   & Both & DA     & Usable   & \miller \\
Praesepe-WDJ084031+214042   & 665139697978259200  & 130.131 & 21.678 & D+S   & Both & DA     & Usable   & \miller \\
Praesepe-WDJ084252+195110   & 661353224747229184  & 130.718 & 19.853 & D+S   & Both & DA     & Usable   & \miller\\
Praesepe-WDJ084321+204330   & 661841163095377024  & 130.841 & 20.725 & S     & Both & DA     & Usable   & \miller \\
Praesepe-WDJ084601+183048   & 660178942032517760  & 131.508 & 18.513 & D     & Both & DA     & Usable   & \miller \\
Praesepe-WDJ085947+422112   & 1008837274655408128 & 134.947 & 42.354 & D+S   & H\&R$^\ddagger$ & DA     & Usable   & \miller \\
Praesepe-WDJ092713+475659   & 825640117568825472  & 141.806 & 47.950 & D+S   & H\&R$^\ddagger$ & DA     & Usable   & \miller \\
RSG5-WDJ200832+444939       & 2082008971824158720 & 302.137 & 44.828 & S     & Both & Peculiar & Usable$^\star$ & \ldots \\
Stock2-WDJ021836+582442     & 458778927573447168  & 34.654  & 58.412 & S     & H\&R & DA     & Usable   & \miller \\
UPK303-WDJ023931+415825     & 337013370615249408  &  39.883 & 41.974 & L     & Both & DA     & Usable   & \miller \\
\bottomrule
\end{tabular}
\begin{minipage}{\linewidth}
\medskip
{\textbf{Notes}. Survey lists the available spectroscopy, with D, L, and S denoting DESI, LAMOST, and SDSS, respectively. Catalog identifies whether the WD is listed as a candidate member in \citet{2024A&A...686A..42H}, the UCC \citep{2023MNRAS.526.4107P}, or both. Host indicates whether the associated cluster provides a sufficiently well-constrained age for analysis. Analysis identifies where the primary WD characterization is presented; DA WDs with usable hosts and spectra from multiple surveys are additionally analyzed here in the cross-survey comparison. Sources with unclassifiable spectra, spectra with insufficient S/N for atmospheric fitting, or excluded host systems are not analyzed in detail. $^\dagger$ denotes a spectrum sufficient for classification but not atmospheric fitting. $^\ddagger$ sources are identified by \citet{2024A&A...686A..42H} as candidate members of HSC~1224, but are retained here as Praesepe members based on their established literature associations and previous analysis in \miller. $^\star$ RSG~5 candidate is excluded based on questionable membership that requires an extreme binary evolutionary channel to be viable.}
\end{minipage}
\end{table*}

We next inspected the Gaia color-magnitude diagrams (CMDs) of the proposed host systems and exclude candidates for which we cannot determine a well-constrained cluster age, either because the member sequence is poorly defined or because the main-sequence turnoff (MSTO) is insufficiently populated. For completeness, we do not enforce a minimum age for potential WD host clusters; the plausibility of the associated WD membership is assessed based on determined WD parameters. Cluster quality criteria exclude WDs associated with CWNU~1149, CWNU~1182, FSR~1017, HSC~190, HSC~453, HSC~913, HSC~1077, HSC~1279, HSC~1542, and HSC~1555. All ten are moving groups, which we retain in the initial search because some have sufficiently well-constrained populations and ages for WD analysis, as demonstrated by the AB~Doradus~Moving~Group and its massive WD GD~50 \citep{2018ApJ...861L..13G}. We also recover ZTF~J2008+4449, a rapidly rotating magnetic WD identified as a candidate RSG~5 member by \miller, \citet{2022AJ....164..215B}, and \citet{2025ApJ...991L...7Y}. \citet{2026A&A...706A.188C} conclude that its age is incompatible with RSG~5 membership, while \citet{2025ApJ...991L...7Y} argued that its kinematics support membership if it formed through a highly accelerated binary-evolution pathway.  Given potential membership requires an extreme binary channel, we exclude it from the sample presented here.

After these exclusions, 44 WDs remain suitable for detailed analysis. Five of these are non-DAs and two are magnetic DAHs. Four of the non-DAs and one DAH have been studied previously, while the DC HSC1224-WDJ092525+525651 and DAH Alessi13-WDJ033301-384025 have not previously been analyzed as cluster WDs. We defer these seven objects to \citet{Miller2026Atypical}, which examines the broader population of magnetic and non-DA cluster WDs. From the remaining 37 non-magnetic DAs, we define two analysis samples with different purposes. The 15 not already included in the recent cluster WD analysis of \miller form the primary sample analyzed in this work. A second sample comprises the 11 non-magnetic DAs with spectra from multiple surveys, which we use to test how the choice of survey spectrum affects the inferred WD and progenitor parameters.


\section{Analysis}
\label{sec:analysis}

Throughout this analysis, we follow the cluster and WD analysis techniques of \miller. We describe the primary fitting procedures below and refer the reader to that work for further details. We apply the same procedure to the 15 DAs in our primary sample and independently to each spectrum in the multi-survey sample. The resulting parameters are presented in Tables~\ref{tab:wds_derived} and \ref{tab:wds_survey_compare}, respectively.

\begin{table*}[htbp]
\footnotesize
\centering
\caption{Adopted atmospheric, cooling, and progenitor parameters for the primary DA WD sample.} 
\label{tab:wds_derived}
\begin{tabular}{cccccccc}
\toprule
Name &
Fit &
$\log g$ &
$T_\textrm{eff}$ &
Radius & 
$M_\textrm{WD}$ &
$t_\textrm{cool}$ & 
$M_\textrm{prog}$ 
\\
&
&
[$\mathrm{cm}\,\mathrm{s}^{-2}$] &
[K] &
[$\mathrm{km}$] &
[$\msun$] & 
[$\mathrm{Myr}$] & 
[$\msun$]
\\
\midrule
Alessi62-WDJ185530+214138 & SDSS & $7.78\pm0.18$ & $57{,}000\pm4{,}000$ & $11{,}500^{+1{,}900}_{-1{,}600}$ & $0.60^{+0.08}_{-0.06}$ & $1.4^{+0.3}_{-0.4}$ & $2.39\pm0.06$ \\
Alessi94-WDJ001937+461007 & SDSS & $8.18\pm0.13$ & $34{,}800\pm1{,}000$ & $8{,}100\pm800$ & $0.76^{+0.08}_{-0.07}$ & $5.9^{+1.2}_{-0.6}$ & $2.87^{+0.05}_{-0.04}$ \\
CWNU1095-WDJ101309+524349 & DESI & $8.31\pm0.05$ & $20{,}300\pm500$ & $7{,}300\pm300$ & $0.81\pm0.03$ & $136^{+15}_{-13}$ & $3.22^{+0.10}_{-0.09}$  \\
HSC381-WDJ175813+172113 & SDSS & $7.96\pm0.06$ & $45{,}000\pm800$ & $9{,}800^{+500}_{-400}$ & $0.66\pm0.03$ & $2.64^{+0.14}_{-0.13}$ & \ldots$^{\rm c}$ \\
HSC448-WDJ214848-035539  & DESI & $8.28\pm0.11$ & $20{,}700\pm1{,}000$ & $7{,}400\pm600$ & $0.79\pm0.07$ & $120\pm30$ & \ldots$^{\rm a}$ \\
HSC448-WDJ222318+000556 & LAMOST & $7.80\pm0.05$ & $56{,}000\pm1{,}000$ & $11{,}300\pm500$ & $0.61\pm0.02$ & $1.46\pm0.07$ & \ldots$^{\rm c}$  \\
HSC1340-WDJ041034+160321 & SDSS & $7.25\pm0.03$ & $23{,}500\pm200$ & $15{,}700^{+400}_{-300}$ & $0.332\pm0.008$ & $8^{+2}_{-3}$ & \ldots$^{\rm b}$ \\
HSC1630-WDJ042914+093722 & LAMOST & $8.54\pm0.08$ & $17{,}900\pm500$ & $6{,}000\pm400$ & $0.95\pm0.05$ & $300^{+50}_{-40}$ & $3.8^{+0.7}_{-0.5}$ \\
Hyades-WDJ040842+292130 & LAMOST & $8.18\pm0.04$ & $7{,}540\pm20$ & $7{,}900\pm200$ & $0.71^{+0.02}_{-0.03}$ & $3{,}950^{+60}_{-100}$ & \ldots$^{\rm a}$  \\
Hyades-WDJ072228+221210 & SDSS & $8.11\pm0.02$ & $16{,}200\pm130$ & $8{,}370\pm120$ & $0.680\pm0.012$ & $203^{+5}_{-9}$ & $3.06^{+0.23}_{-0.18}$  \\
Melotte111-WDJ121100+225220 & LAMOST & $7.94\pm0.04$ & $19{,}300\pm200$ & $9{,}500\pm300$ & $0.59\pm0.02$ & $74\pm5$ & $2.77^{+0.08}_{-0.02}$  \\
M67-WDJ085028+105417 & SDSS & $8.12\pm0.09$ & $25{,}600\pm700$ & $8{,}400^{+600}_{-500}$ & $0.71\pm0.05$ & $25^{+10}_{-6}$ & $1.343^{+0.010}_{-0.030}$ \\
M67-WDJ085119+114840 & SDSS &$7.37\pm0.12$ & $61{,}000\pm3{,}000$ & $16{,}300^{+1{,}600}_{-1{,}500}$ & $0.47^{+0.04}_{-0.03}$ & $0.03^{+0.30}_{-0.03}$ & \ldots$^{\rm b}$ \\
M67-WDJ085425+113829 & DESI & $8.04\pm0.04$ & $26{,}500\pm300$ & $8{,}900\pm300$ & $0.66\pm0.02$ & $16.9^{+2.3}_{-0.9}$ & $1.343^{+0.010}_{-0.031}$  \\
Praesepe-WDJ082839+194343 & SDSS & $8.31\pm0.03$ & $15{,}700\pm300$ & $7{,}240\pm160$ & $0.805\pm0.019$ & $299^{+21}_{-13}$ & $3.04^{+0.09}_{-0.16}$ \\
\bottomrule
\end{tabular}
\begin{minipage}{\linewidth}
\medskip
{\textbf{Notes}. Progenitor masses are reported only for WDs retained as plausible single-star cluster members. $^{a}$ indicates that the WD cooling age exceeds the cluster age; $^{b}$ indicates that the WD mass is too low to have formed through single-star evolution within a Hubble time; and $^{c}$ indicates that the formal progenitor mass is extremely inconsistent with the observed WD mass under single-star evolution.}
\end{minipage}
\end{table*}

\begin{table*}[htbp]
\tiny
\centering
\caption{Cooling and progenitor parameters from independent fits to non-magnetic DA WDs observed by multiple surveys.}
\label{tab:wds_survey_compare}
\begin{tabular}{cccccccccc}
\toprule
Name &
\multicolumn{3}{c}{$M_\textrm{WD}$ [$\msun$]} &
\multicolumn{3}{c}{$t_\textrm{cool}$ [$\mathrm{Myr}$]} &
\multicolumn{3}{c}{$M_\textrm{prog}$ [$\msun$]}
\\
\cmidrule(lr){2-4}
\cmidrule(lr){5-7}
\cmidrule(lr){8-10}
&
DESI &
LAMOST &
SDSS &
DESI &
LAMOST &
SDSS &
DESI &
LAMOST &
SDSS
\\
\midrule
HSC448-WDJ222318+000556 & $0.56^{+0.05}_{-0.04}$ & $0.608^{+0.021}_{-0.020}$ & \ldots & $1.0^{+0.2}_{-0.5}$ & $1.46\pm0.07$ & \ldots & \ldots & \ldots & \ldots \\
Hyades-WDJ072228+221210 & \ldots & $0.65\pm0.02$ & $0.680\pm0.012$ & \ldots & $153^{+20}_{-9}$ & $203^{+5}_{-9}$ & \ldots & $2.95^{+0.20}_{-0.15}$ & $3.06^{+0.23}_{-0.18}$ \\
M67-WDJ085425+113829 & $0.66\pm0.02$ & \ldots & $0.77\pm0.06$ & $16.9^{+2.3}_{-0.9}$ & \ldots & $30^{+14}_{-9}$ & $1.343^{+0.010}_{-0.031}$ & \ldots & $1.344^{+0.010}_{-0.031}$ \\
Praesepe-WDJ082839+194343 & $0.82\pm0.03$ & \ldots & $0.805\pm0.019$ & $276\pm19$ & \ldots & $299^{+21}_{-13}$ & $2.98^{+0.08}_{-0.15}$ & \ldots & $3.04^{+0.09}_{-0.16}$ \\
Praesepe-WDJ083947+194611 & $0.883\pm0.013$ & \ldots & $0.68\pm0.04$ & $353^{+11}_{-13}$ & \ldots & $130^{+16}_{-21}$ & $3.18^{+0.10}_{-0.19}$ & \ldots & $2.72^{+0.05}_{-0.10}$ \\
Praesepe-WDJ084013+184326 & $0.92\pm0.04$ & $0.791\pm0.019$ & $0.82\pm0.03$ & $290\pm30$ & $335^{+20}_{-19}$ & $288^{+18}_{-19}$ & $3.01^{+0.10}_{-0.16}$ & $3.13^{+0.10}_{-0.18}$ & $3.01^{+0.09}_{-0.15}$ \\
Praesepe-WDJ084028+194334 & \ldots & $0.67\pm0.05$ & $0.698\pm0.012$ & \ldots & $82^{+18}_{-13}$ & $116^{+7}_{-3}$ & \ldots & $2.65^{+0.05}_{-0.09}$ & $2.70^{+0.05}_{-0.10}$ \\
Praesepe-WDJ084031+214042 & $0.854\pm0.013$ & \ldots & $0.828\pm0.013$ & $225^{+10}_{-6}$ & \ldots & $225^{+8}_{-9}$ & $2.88^{+0.06}_{-0.13}$ & \ldots & $2.88^{+0.06}_{-0.13}$ \\
Praesepe-WDJ084252+195110 & $0.795\pm0.013$ & \ldots & $0.744\pm0.013$ & $217^{+7}_{-5}$ & \ldots & $203^{+5}_{-7}$ & $2.86^{+0.06}_{-0.12}$ & \ldots & $2.84^{+0.06}_{-0.12}$ \\
Praesepe-WDJ085947+422112 & $0.94^{+0.02}_{-0.03}$ & \ldots & $0.851\pm0.019$ & $420^{+40}_{-30}$ & \ldots & $312^{+21}_{-16}$ & $3.4^{+0.2}_{-0.3}$ & \ldots & $3.07^{+0.10}_{-0.16}$ \\
Praesepe-WDJ092713+475659 & $0.895\pm0.019$ & \ldots & $0.900\pm0.013$ & $430\pm30$ & \ldots & $480\pm20$ & $3.43^{+0.18}_{-0.26}$ & \ldots & $3.7^{+0.2}_{-0.3}$ \\
\bottomrule
\end{tabular}
\begin{minipage}{\linewidth}
\medskip
{\textbf{Notes}. HSC448-WDJ222318+000556 is included for completeness but excluded from the analysis in Section~\ref{subsec:cross_survey} because its inferred progenitor masses are grossly inconsistent with its WD mass under single-star evolution.}
\end{minipage}
\end{table*}


\subsection{Spectral Fitting and WD Parameters}

\begin{figure*}[ht]
    \centering
    \includegraphics[width=0.98\textwidth]{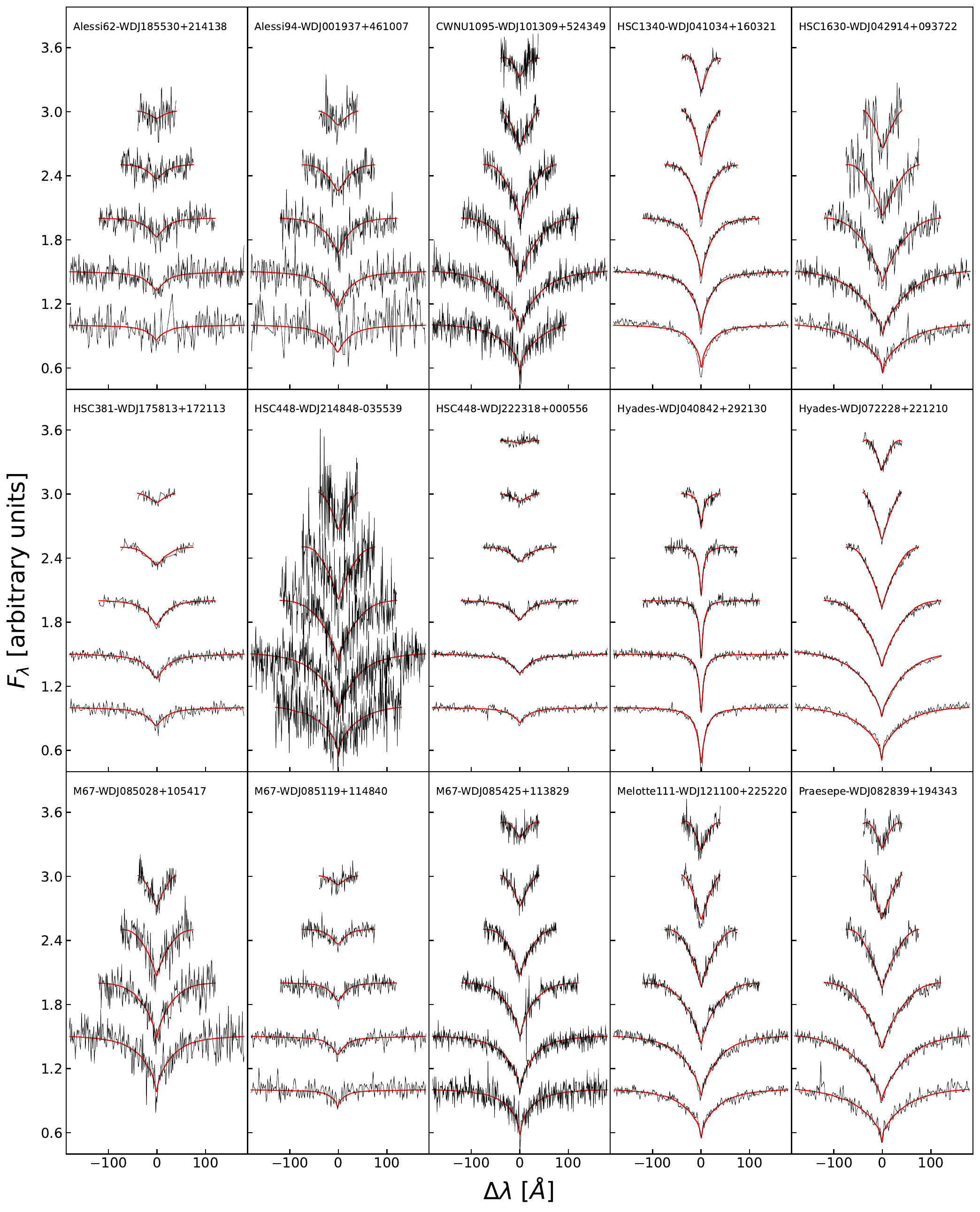}
    \caption{Normalized Balmer-line profiles for the 15 DA WDs in the primary sample, with the best-fit atmosphere models overlaid. Balmer lines are vertically offset from H$\alpha$ through H$\zeta$, as available for each spectrum.} 
    \label{fig:new_fits} 
\end{figure*}

We determine atmospheric parameters for the non-magnetic DA WDs using the NLTE pure-H atmosphere models of \citet{2011ApJ...730..128T}, following the general Balmer-line fitting procedure of \citet{2005ApJS..156...47L}. We first fit each observed spectrum with a grid of synthetic models together with a polynomial of up to ninth order in wavelength to account for residual continuum calibration differences. The spectrum is then normalized using fixed continuum points outside the Balmer lines, and the normalized line profiles are fit simultaneously for $T_{\rm eff}$ and $\log g$ using Levenberg--Marquardt nonlinear least-squares minimization. For DAs with $T_{\rm eff}>40{,}000$ K, we instead fit the metal-enriched models of \citet{2010ApJ...720..581G} to account for the Balmer-line problem in hot WDs, although the resulting differences are generally minimal until substantially higher temperatures. The resulting Balmer-line fits for the 15 DA WDs in the primary sample are shown in Fig.~\ref{fig:new_fits}. For the multi-survey sample, we use the same set of Balmer lines for each spectrum of a given WD, restricted to those reliably measurable in all available spectra, so that the comparison is not driven by differences in line selection. In all cases, we use the thick-H envelope cooling models of \citet{2020ApJ...901...93B} to map the fitted $T_{\rm eff}$ and $\log g$ to WD mass, radius, and cooling age. 

\subsection{Cluster and Progenitor Parameters}

\begin{figure*}[ht]
    \centering
    \includegraphics[width=0.9\textwidth]{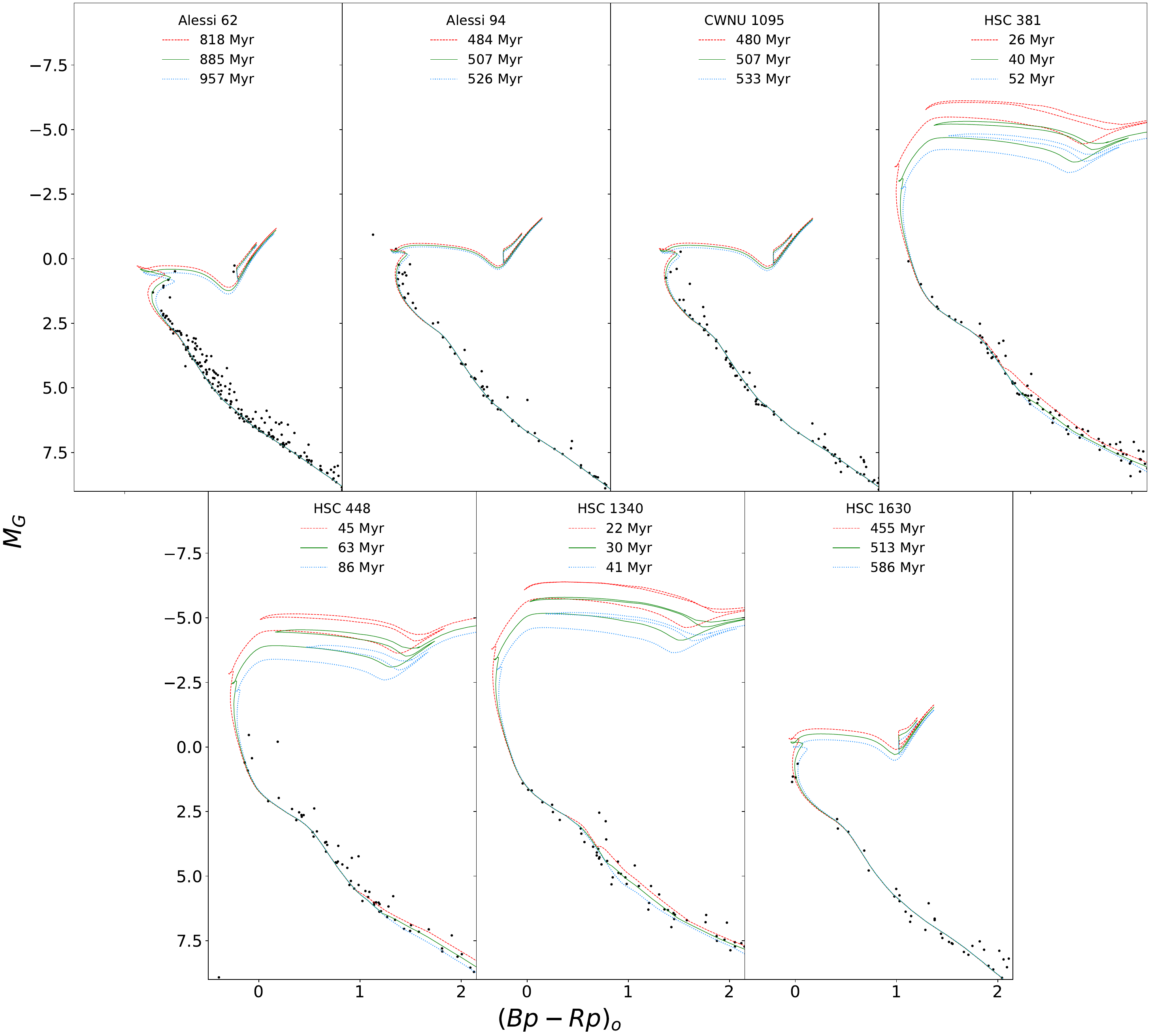}
    \caption{CMDs and isochrone fits for the seven clusters analyzed in this work. Green shows the best-fit PARSEC isochrone, with the red and blue isochrones indicating the lower and upper age limits, respectively.} 
    \label{fig:cluster_ages} 
\end{figure*}

Cluster parameters are obtained using the PARSEC \citep{2012MNRAS.427..127B,2014MNRAS.445.4287T,2014MNRAS.444.2525C,2018MNRAS.476..496F,2022A&A...665A.126N} isochrone-fitting procedure developed by \miller, which emphasizes the MSTO while also considering the broader main-sequence and giant branch morphology where available; we refer the reader to \miller for full details of the fitting procedure. We adopt previously determined cluster parameters from \miller where available and apply the same procedure to clusters without fits from those works. We determine the best-fit age using these CMD features, using a weighted $\chi^2$ calculation to estimate the age bounds. The new cluster fits are shown in Fig.~\ref{fig:cluster_ages}, and the adopted parameters for all host clusters, including both newly fitted and previously determined values, are summarized in Table~\ref{tab:clusters}. The quoted age uncertainties are from the fitting procedure and do not include additional systematics from evolutionary models or unresolved binaries.

\begin{table}[ht]
\tiny
\caption{Adopted properties of clusters represented in the primary and multi-survey analysis samples.}
\label{tab:clusters}
\centering
\setlength{\tabcolsep}{3pt}
\begin{tabular}{cccccc}
\toprule
Name &
Distance &
$A_\mathrm{V}$ &
$Z$ &
Age &
Source
\\
&
[pc] &
[mag] &
&
[Myr] &
\\
\midrule
Alessi 62    & 606 & 0.565 & 0.018 & $885^{+72}_{-67}$ & this work \\
Alessi 94    & 553 & 0.159 & 0.016 & $507^{+19}_{-23}$ & this work \\
CWNU 1095    & 245 & 0.053 & 0.018 & $507^{+26}_{-27}$  & this work \\
HSC 381      & 313 & 0.181 & 0.017 & $40^{+12}_{-14}$  & this work \\
HSC 448      & 344 & 0.116 & 0.014 & $63^{+23}_{-18}$  & this work \\
HSC 1340     & 120 & 0.120 & 0.020 & $30^{+11}_{-8}$   & this work \\
HSC 1630     & 134 & 0.027 & 0.010 & $513^{+73}_{-58}$ & this work \\
Hyades       & 47  & 0.000 & 0.024 & $646^{+85}_{-84}$       & \miller \\
Melotte 111  & 85  & 0.000 & 0.014 & $617^{+11}_{-42}$ & \miller \\
M67          & 837 & 0.132 & 0.012 & $3{,}819^{+292}_{-86}$  & \miller \\
Praesepe     & 183 & 0.059 & 0.019 & $741^{+72}_{-30}$       & \miller  \\
\bottomrule
\end{tabular}
\begin{minipage}{\linewidth}
\medskip
{\textbf{Notes}. Distance from \citet{2024A&A...686A..42H}, other parameters determined using the  heuristic isochrone fitting technique of \miller, with the source of the adopted fit listed in the final column.}
\end{minipage}
\end{table}

For each WD, we subtract the WD cooling age from the adopted cluster age to obtain the progenitor lifetime and use PARSEC isochrones at the adopted cluster metallicity to infer the corresponding progenitor mass. As in \miller, we determine the progenitor mass from the lowest-mass star to have reached the onset of the asymptotic giant branch. We report progenitor masses only when the resulting evolutionary timescale and WD parameters are consistent with a single-star cluster origin.


\section{Results and Discussion}
\label{sec:results}

We focus on two complementary uses of the archival spectroscopic census: identifying and characterizing new cluster WDs, and testing the consistency of WD parameters derived from different spectroscopic surveys. We first assess the 15 non-magnetic DA WDs in the primary sample, including their consistency with single-star cluster membership. We then examine the locations of the retained WDs in the empirical IFMR and quantify their consistency with the \miller relation. Finally, we use WDs observed by multiple surveys to compare the inferred atmospheric, cooling, and progenitor properties and assess the consistency of the resulting IFMR measurements.


\subsection{Newly Characterized Cluster WDs}
\label{subsec:new_wds}

Of the 15 WDs in the primary DA sample, nine are consistent with single-star cluster membership and are retained for further analysis. We reject the remaining six candidates from the single-star cluster WD sample. HSC448-WDJ214848-035539 and Hyades-WDJ040842+292130 have cooling ages that far exceed the ages of their proposed host clusters, ruling out membership. HSC1340-WDJ041034+160321 and M67-WDJ085119+114840 instead have WD masses of only $0.33$ and $0.47\msun$, respectively, too low to have formed through single-star evolution within a Hubble time. Finally, HSC381-WDJ175813+172113 and HSC448-WDJ222318+000556 have cooling ages shorter than their host ages, but imply progenitor masses of approximately $8.2$ and $6.5\msun$ despite WD masses of only $0.66$ and $0.61\msun$, respectively. The implied progenitor and WD mass combinations for these two objects are grossly inconsistent with single-star evolution, ruling out a single-star cluster origin. We therefore do not report progenitor masses for any of these six sources in Table~\ref{tab:wds_derived}.

The six rejected candidates are not unexpected given the intentionally broad initial selection. Four of the 15 WDs in our primary sample are associated with the very young HSC~381, HSC~448 and HSC~1340 clusters, where only particularly massive progenitors could have had time to form WDs through single-star evolution. For completeness, we did not limit candidates using Gaia photometric masses or cooling-age estimates, as is typical when selecting targets for spectroscopic follow-up in the literature. Because our sample is drawn entirely from archival spectra, there was no need to pre-screen candidates to avoid spending valuable telescope time on objects whose photometric WD parameters make cluster membership unlikely. Of the 22 targets followed up spectroscopically by \miller, 19 were consistent with single-star cluster membership, including 18 of 19 non-magnetic DAs. The higher contamination rate in our sample is thus expected from the deliberately broad initial selection.

Hyades-WDJ072228+221210 is notable as a likely Hyades tidal-tail member, as classified by \citet{2026arXiv260711299J}. Although it lies well outside the compact spatial distribution of the cluster, this is consistent with the highly extended nature of the Hyades tidal tails. The spectroscopic parameters derived here provide an independent check on this association: its WD mass and cooling age are consistent with a progenitor lifetime and initial mass expected for single-star evolution at the Hyades age. The Hyades is the nearest open cluster and one of the most extensively studied, yet has long shown a substantial deficit of WDs relative to expectations \citep[e.g.,][]{1992AJ....104.1876W}. This deficit remains apparent in modern dynamical modeling: the model of \citet{2026arXiv260711299J} that reproduces the Hyades core population after accounting for unresolved binaries predicts approximately 40 WDs in the core at the cluster age, compared with only eight observed. Along with the ultramassive escaped member identified by \citet{2023ApJ...956L..41M}, this tidal tail WD provides further evidence that dynamical dispersal has contributed to the long-standing Hyades WD deficit.


\subsection{IFMR Consistency}
\label{subsec:ifmr}

\begin{figure*}[ht]
    \centering
    \includegraphics[width=\textwidth]{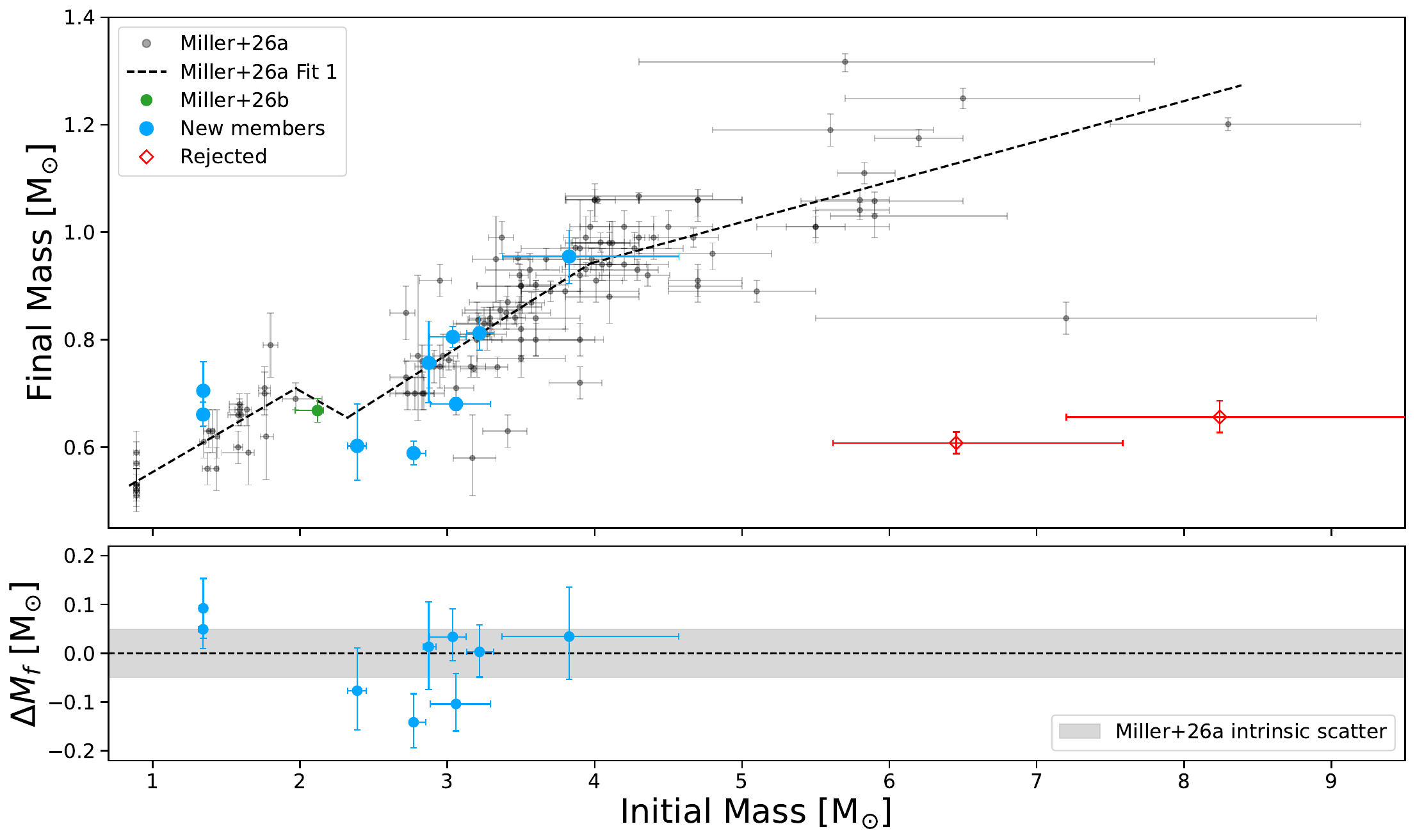}
    \caption{Top: cluster IFMR including the nine newly retained WDs from this work. Black points show the DA sample from \miller, with its best-fit relation shown in black, while the additional cluster DA WD from \citet{2026ApJ..1007..183M} is shown in green. New members are shown in blue, while the two rejected candidates with formal progenitor mass solutions are shown in red. Bottom: residuals of the nine retained WDs from the \miller relation; the shaded region marks the intrinsic scatter from that work.}
    \label{fig:IFMR_new}
\end{figure*}

Together with the 122 DA cluster WDs analyzed by \miller and the additional cluster DA WD identified in \citet{2026ApJ..1007..183M}, the nine newly retained WDs bring the combined IFMR sample to 132 objects. The existing sample is available at wdifmr.org\footnote{\url{https://wdifmr.org}} with the additional WDs presented here to be added upon acceptance of this manuscript. We quantify the consistency of the nine newly retained WDs with the best-fit relation of \miller using 100,000 Monte Carlo realizations. In each realization, we draw the WD and progenitor masses from their asymmetric uncertainties and sample the published IFMR uncertainties, using the same IFMR realization for all nine WDs in each trial. For each object, we then calculate the final-mass residual, $\Delta M_f=M_f-M_{f,\mathrm{IFMR}}$. We assess the significance of these residuals by combining their Monte Carlo uncertainties with the intrinsic scatter of $\sigma_{\rm int}=0.049\msun$ measured by \miller. The resulting IFMR comparison is shown in Fig.~\ref{fig:IFMR_new}, with the residuals shown in the bottom panel.

Six of the nine WDs lie within $1\sigma$ of the \miller best-fit relation, and all nine lie within $2\sigma$. Melotte111-WDJ121100+225220 shows the largest offset, $\Delta M_f=-0.142^{+0.059}_{-0.053}\msun$, corresponding to a $1.9\sigma$ deviation. The offset is modest given the measurement uncertainties, systematics, and the poorly constrained behavior of the IFMR between $M_{\rm prog}\sim2$--$2.7\msun$, as discussed in \miller and \citet{2026ApJ..1007..183M}. We therefore retain it in the single-star sample while noting that its IFMR position may indicate a greater likelihood of a binary evolutionary history, since binary interaction can alter the remnant mass relative to the single-star expectation \citep{2000MNRAS.319..215H,2008AJ....135.2163R}. More generally, binary evolution products, including WDs descended from blue stragglers, may yield progenitor masses that are inconsistent with those inferred by assuming single-star evolution from the cluster age and WD cooling age. Overall, the nine retained WDs show no evidence for a change to the current empirical IFMR. Given this consistency, we do not derive an updated relation here.


\subsection{Cross-Survey Spectroscopic Consistency}
\label{subsec:cross_survey}

WDs observed by multiple surveys allow a direct comparison of parameters derived from independent spectra. For cluster WDs, the cluster constraints provide additional leverage unavailable for field WDs, allowing differences in the atmospheric parameters to propagate to progenitor properties. 11 non-magnetic DA WDs in our sample have spectra from multiple surveys (Table~\ref{tab:wds_survey_compare}). Ten have parameters consistent with a single-star origin in their respective host clusters; the exception is HSC448-WDJ222318+000556. We use these ten WDs to compare the atmospheric parameters, WD masses, cooling ages, and progenitor masses derived from each survey spectrum. For each WD, we fit the same six Balmer lines in all available spectra, corresponding to the largest number of lines that could be measured reliable across the majority of spectra included in the comparison. This avoids differences in the fits driven by line selection, which has been shown to significantly affect the inferred atmospheric parameters \citep{2019ApJ...885...86S}. For each spectrum, we estimate the S/N as the median per-pixel flux divided by its uncertainty over 5100--5500~\AA.

\begin{figure*}[ht]
    \centering
    \includegraphics[width=\textwidth]{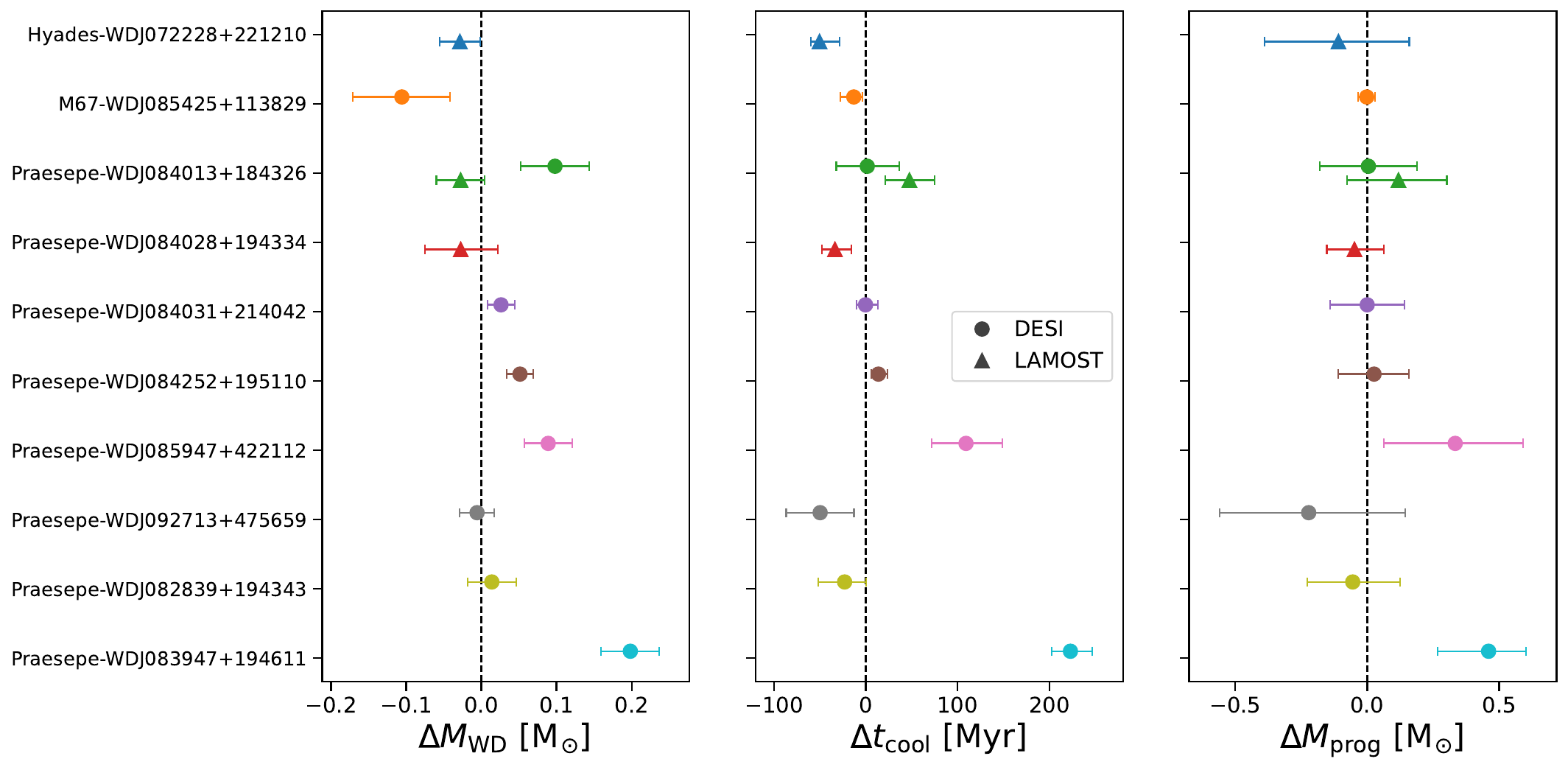}
    \caption{Differences in WD mass, cooling age, and progenitor mass derived from independent DESI and LAMOST spectra relative to the corresponding SDSS solutions for the ten WDs retained in the cross-survey analysis. Differences are defined as $\Delta X=X_{\rm survey}-X_{\rm SDSS}$. Colors identify individual WDs and symbols distinguish DESI and LAMOST measurements.}
    \label{fig:cross_survey_compare}
\end{figure*}

The comparison sample comprises nine WDs with spectra from two surveys, while Praesepe-WDJ084013+184326 has spectra from all three. Eight of the WDs are in Praesepe, making them especially valuable because differences in the inferred progenitor properties arise purely from the spectroscopic solutions. Fig.~\ref{fig:cross_survey_compare} compares the differences in WD mass, cooling age, and progenitor mass derived from DESI and LAMOST spectra relative to the corresponding SDSS solutions, which provide a common reference for all ten WDs. The Praesepe comparisons show a clear tendency for the DESI solutions to yield higher WD masses than SDSS, while the M67 WD shows the opposite behavior.

For a uniform quantitative comparison, we focus on the seven Praesepe WDs observed by both DESI and SDSS. DESI yields a higher WD mass in six of the seven cases, with a mean DESI--SDSS offset of $+0.067\msun$ and a standard deviation of $0.069\msun$. Changing the adopted spectral resolution by a factor of two had a negligible effect on the mean DESI--SDSS mass offset, suggesting that the assumed resolution does not account for the discrepancy. The inferred progenitor masses differ by as much as $0.459\msun$ between the two surveys. These differences are not captured by the WD masses alone. For example, Praesepe-WDJ092713+475659 has nearly identical DESI and SDSS WD masses, but its inferred progenitor masses differ by approximately $0.3\msun$, largely because the DESI fit is $\approx600$~K hotter. This illustrates that agreement in WD mass does not necessarily imply agreement in the broader evolutionary parameters, as differences in $T_{\rm eff}$ can still substantially alter the cooling age and inferred progenitor mass.

The systematic displacement of the DESI solutions motivates an examination of its implications for the empirical IFMR. This comparison does not require repeat spectroscopy, so we consider all ten single-star DA WDs in our sample with DESI spectra. These comprise eight WDs from the multi-survey sample, three from the primary sample, with two WDs common to both, together with Praesepe-WDJ084601+183048 which was previously analyzed by \miller. None of the spectroscopy used to construct the \miller IFMR came from DESI, so the published relation provides a spectroscopically independent benchmark for the DESI-derived parameters. For the additional Praesepe WD, we derive $M_{\rm WD}=0.915\pm0.019\msun$, and $M_{\rm prog}=3.24^{+0.13}_{-0.20}\msun$ from its DESI spectrum. 

\begin{figure}[ht]
    \centering
    \includegraphics[width=\columnwidth]{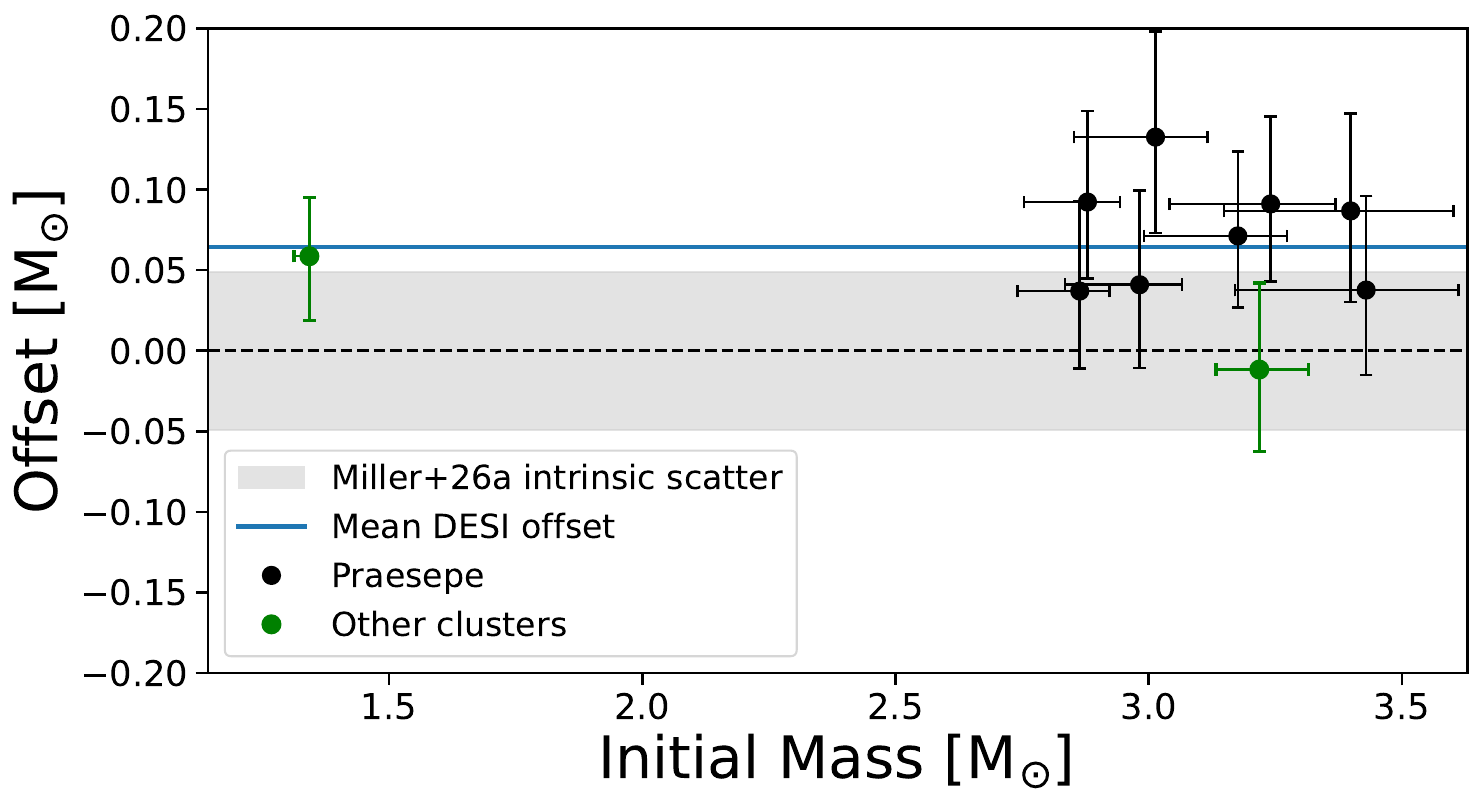}
    \caption{IFMR residuals for the ten single-star DA WDs with DESI spectra, measured relative to the best-fit IFMR of \miller. The shaded region shows the intrinsic scatter of $\sigma_{\rm int}=0.049\msun$, and the blue line marks the mean DESI offset. Praesepe WDs are shown in black and WDs from other clusters in green.}
    \label{fig:IFMR_DESI_residuals}
\end{figure}

We quantify the displacement of the DESI-derived parameters from the best-fit relation of \miller using the same Monte Carlo framework as in Section~\ref{subsec:ifmr}. We generate 100,000 realizations by drawing the WD and progenitor masses from their asymmetric uncertainties and sampling the published IFMR uncertainties, applying the same IFMR realization for all ten WDs in each trial. For each WD, we calculate the vertical residual, $\Delta M_f=M_f-M_{f,\mathrm{IFMR}}$. Rather than assessing each WD individually, we use the mean residual of the ten DESI solutions to test for a systematic offset. We account for the $\sigma_{\rm int}=0.049\msun$ intrinsic scatter measured by \miller by including an independent draw from this scatter for each WD before calculating the sample mean in each realization. The resulting individual residuals and mean offset are shown in Fig.~\ref{fig:IFMR_DESI_residuals}.

Nine of the ten DESI solutions lie above the \miller relation, and we find a mean residual of $+0.06\pm0.04\msun$. The magnitude of this offset is comparable to the $0.05$--$0.06\msun$ excess of DESI spectroscopic over photometric DA masses reported by \citet{2026ApJ..1000..216K}, which they attributed to problems with the broad hydrogen-line profiles in DESI spectra. Together with the Praesepe comparison, this suggests a systematic difference in the DESI spectroscopic solutions rather than an offset driven by the adopted cluster properties. The resulting shifts in WD masses, cooling ages, and progenitor masses demonstrate that systematic differences between spectroscopic surveys can affect studies that combine heterogeneous WD spectroscopy, including empirical IFMR measurements.


\section{Summary and Conclusions}
\label{sec:conclusions}

This work presents a spectroscopic census of candidate cluster WDs from DESI, LAMOST, and SDSS. By crossmatching cluster memberships with archival spectroscopy from the three surveys, we identify 62 candidate cluster WDs with available spectra. After accounting for spectral quality and the suitability of the proposed host clusters, 44 WDs remain for detailed analysis. Seven are magnetic or non-DA WDs examined separately in \citet{Miller2026Atypical}, leaving 37 non-magnetic DAs. Within this DA sample, 15 WDs not included in the recent cluster WD study of \miller form our primary sample, while 11 WDs have spectra from multiple surveys, ten of which are retained for a cross-survey comparison.

Nine of the 15 WDs in the primary sample have properties consistent with single-star membership in their proposed host clusters, while the remaining six are excluded from the single-star cluster sample. The nine retained WDs expand the spectroscopically characterized cluster WD sample and, together with the additional cluster DA WD identified in \citet{2026ApJ..1007..183M}, bring the combined cluster DA IFMR sample to 132 objects. All nine lie within $2\sigma$, and six of nine within $1\sigma$, of the \miller best-fit relation, suggesting these additions do not warrant a change to the empirical IFMR. Notably, one member, Hyades-WDJ072228+221210, is likely a Hyades tidal-tail member, adding to the evidence that dynamical dispersal has contributed to the long-standing deficit of WDs associated with the cluster.

Among the seven Praesepe WDs observed by both DESI and SDSS, DESI yields higher WD masses in six cases, with a mean difference of $+0.067\msun$, while the inferred progenitor masses differ by as much as $0.459\msun$. A broader test using all ten single-star DA WDs with DESI spectra finds that nine lie above the \miller IFMR, with a mean residual of $+0.06\pm0.04\msun$. The magnitude of this offset is similar to the $0.05$--$0.06\msun$ excess of DESI spectroscopic over photometric DA masses reported by \citet{2026ApJ..1000..216K}. The cross-survey sample remains relatively small and is dominated by Praesepe WDs, and larger samples will be needed to determine how broadly the measured offsets apply. 

Together, these comparisons indicate that systematic differences between surveys can affect WD masses, cooling ages, and inferred progenitor masses. Studies combining heterogeneous WD spectroscopy, particularly empirical IFMR analyses, may therefore benefit from cross-survey calibration or an explicit treatment of the associated systematic uncertainties.

Wide-field spectroscopic surveys provide an increasingly powerful way to grow the cluster WD sample without dedicated follow-up. Future releases from DESI, SDSS, and LAMOST will continue to increase the available archival spectroscopy, while upcoming public data from WEAVE \citep{2024MNRAS.530.2688J} and 4MOST \citep{2019Msngr.175....3D} will add substantial new spectroscopic samples. Together, these surveys should substantially increase both the number of spectroscopically characterized cluster WDs and the overlap between independent datasets, allowing survey-dependent systematics to be better quantified. Combined with the improved membership constraints expected from Gaia DR4, these data will provide larger, better characterized samples for studying WD evolution, cluster populations, and the IFMR.


\begin{acknowledgments}

D.R.M. and J.H. acknowledge support from the Natural Sciences and Engineering Research Council of Canada Discovery Grants DG-RGPIN-2022-03051 and DG-RGPIN-2023-04486. C.L. and H.Y. acknowledge support from the National Natural Science Foundation of China (NSFC) through grants 12233013 and CMS-CSST-2025-A13. P.-E.T. acknowledges funding from the European Research Council under the European Union's Horizon 2020 research and innovation program number 101002408 (MOS100PC).

This work has made use of data from the European Space Agency (ESA) mission {\it Gaia} (\url{https://www.cosmos.esa.int/gaia)}, processed by the {\it Gaia} Data Processing and Analysis Consortium (DPAC, \url{https://www.cosmos.esa.int/web/gaia/dpac/consortium}). Funding for the DPAC has been provided by national institutions, in particular the institutions participating in the {\it Gaia} Multilateral Agreement. 

Funding for the Sloan Digital Sky Survey V has been provided by the Alfred P. Sloan Foundation, the Heising-Simons Foundation, the National Science Foundation, and the Participating Institutions. SDSS acknowledges support and resources from the Center for High-Performance Computing at the University of Utah. SDSS telescopes are located at Apache Point Observatory, funded by the Astrophysical Research Consortium and operated by New Mexico State University, and at Las Campanas Observatory, operated by the Carnegie Institution for Science. The SDSS web site is \url{www.sdss.org}. SDSS is managed by the Astrophysical Research Consortium for the Participating Institutions of the SDSS Collaboration, including the Carnegie Institution for Science, Chilean National Time Allocation Committee (CNTAC) ratified researchers, Caltech, the Gotham Participation Group, Harvard University, Heidelberg University, The Flatiron Institute, The Johns Hopkins University, L'Ecole polytechnique f\'{e}d\'{e}rale de Lausanne (EPFL), Leibniz-Institut f\"{u}r Astrophysik Potsdam (AIP), Max-Planck-Institut f\"{u}r Astronomie (MPIA Heidelberg), Max-Planck-Institut f\"{u}r Extraterrestrische Physik (MPE), Nanjing University, National Astronomical Observatories of China (NAOC), New Mexico State University, The Ohio State University, Pennsylvania State University, Smithsonian Astrophysical Observatory, Space Telescope Science Institute (STScI), the Stellar Astrophysics Participation Group, Universidad Nacional Aut\'{o}noma de M\'{e}xico, University of Arizona, University of Colorado Boulder, University of Illinois at Urbana-Champaign, University of Toronto, University of Utah, University of Virginia, Yale University, and Yunnan University. 

This research used data obtained with the Dark Energy Spectroscopic Instrument (DESI). DESI construction and operations is managed by the Lawrence Berkeley National Laboratory. This material is based upon work supported by the \href{https://www.energy.gov}{U.S. Department of Energy}, Office of Science, Office of High-Energy Physics, under Contract No. DE--AC02--05CH11231, and by the National Energy Research Scientific Computing Center, a DOE Office of Science User Facility under the same contract. Additional support for DESI was provided by the \href{https://www.nsf.gov}{U.S. National Science Foundation} (NSF), Division of Astronomical Sciences under Contract No. AST-0950945 to the NSF's National Optical-Infrared Astronomy Research Laboratory; the \href{https://stfc.ukri.org}{Science and Technology Facilities Council of the United Kingdom}; the \href{https://www.moore.org}{Gordon and Betty Moore Foundation}; the \href{https://www.hsfoundation.org}{Heising-Simons Foundation}; the \href{https://www.cea.fr}{French Alternative Energies and Atomic Energy Commission} (CEA); the \href{https://secihti.mx/}{National Council of Humanities, Science and Technology of Mexico} (CONAHCYT); the \href{http://www.mineco.gob.es}{Ministry of Science and Innovation of Spain} (MICINN), and by the DESI Member Institutions: \url{https://www.desi.lbl.gov/collaborating-institutions}. The DESI collaboration is honored to be permitted to conduct scientific research on I'oligam Du'ag (Kitt Peak), a mountain with particular significance to the \href{https://www.tonation-nsn.gov/}{Tohono O'odham Nation}. Any opinions, findings, and conclusions or recommendations expressed in this material are those of the author(s) and do not necessarily reflect the views of the U.S. National Science Foundation, the U.S. Department of Energy, or any of the listed funding agencies. 

This work made use of data from LAMOST (Large Sky Area Multi-Object Fiber Spectroscopic Telescope, also known as the Guoshoujing Telescope; \url{https://cstr.cn/31118.02.LAMOST}). LAMOST is a Chinese national mega-science facility operated by the National Astronomical Observatories, Chinese Academy of Sciences.
\end{acknowledgments}

\facilities{Gaia, Mayall (DESI), LAMOST, Sloan}

\software{Astropy \citep{2013A&A...558A..33A,2018AJ....156..123A,2022ApJ...935..167A}}

\section*{Data Availability} Much of the single-star cluster WD data and combined IFMR sample presented in this work is available at
\url{https://wdifmr.org}. The additional data introduced in this work will be made available there upon acceptance of this manuscript.


\bibliographystyle{aasjournalv7.1}
\bibliography{main}

@ARTICLE{2016A&A...595A...1G,
       author = {{Gaia Collaboration} and {Prusti}, T. and {de Bruijne}, J.~H.~J. and {Brown}, A.~G.~A. and {Vallenari}, A. and {Babusiaux}, C. and {Bailer-Jones}, C.~A.~L. and {Bastian}, U. and {Biermann}, M. and {Evans}, D.~W. and {Eyer}, L. and {Jansen}, F. and {Jordi}, C. and {Klioner}, S.~A. and {Lammers}, U. and {Lindegren}, L. and {Luri}, X. and {Mignard}, F. and {Milligan}, D.~J. and {Panem}, C. and {Poinsignon}, V. and {Pourbaix}, D. and {Randich}, S. and {Sarri}, G. and {Sartoretti}, P. and {Siddiqui}, H.~I. and {Soubiran}, C. and {Valette}, V. and {van Leeuwen}, F. and {Walton}, N.~A. and {Aerts}, C. and {Arenou}, F. and {Cropper}, M. and {Drimmel}, R. and {H{\o}g}, E. and {Katz}, D. and {Lattanzi}, M.~G. and {O'Mullane}, W. and {Grebel}, E.~K. and {Holland}, A.~D. and {Huc}, C. and {Passot}, X. and {Bramante}, L. and {Cacciari}, C. and {Casta{\~n}eda}, J. and {Chaoul}, L. and {Cheek}, N. and {De Angeli}, F. and {Fabricius}, C. and {Guerra}, R. and {Hern{\'a}ndez}, J. and {Jean-Antoine-Piccolo}, A. and {Masana}, E. and {Messineo}, R. and {Mowlavi}, N. and {Nienartowicz}, K. and {Ord{\'o}{\~n}ez-Blanco}, D. and {Panuzzo}, P. and {Portell}, J. and {Richards}, P.~J. and {Riello}, M. and {Seabroke}, G.~M. and {Tanga}, P. and {Th{\'e}venin}, F. and {Torra}, J. and {Els}, S.~G. and {Gracia-Abril}, G. and {Comoretto}, G. and {Garcia-Reinaldos}, M. and {Lock}, T. and {Mercier}, E. and {Altmann}, M. and {Andrae}, R. and {Astraatmadja}, T.~L. and {Bellas-Velidis}, I. and {Benson}, K. and {Berthier}, J. and {Blomme}, R. and {Busso}, G. and {Carry}, B. and {Cellino}, A. and {Clementini}, G. and {Cowell}, S. and {Creevey}, O. and {Cuypers}, J. and {Davidson}, M. and {De Ridder}, J. and {de Torres}, A. and {Delchambre}, L. and {Dell'Oro}, A. and {Ducourant}, C. and {Fr{\'e}mat}, Y. and {Garc{\'\i}a-Torres}, M. and {Gosset}, E. and {Halbwachs}, J. -L. and {Hambly}, N.~C. and {Harrison}, D.~L. and {Hauser}, M. and {Hestroffer}, D. and {Hodgkin}, S.~T. and {Huckle}, H.~E. and {Hutton}, A. and {Jasniewicz}, G. and {Jordan}, S. and {Kontizas}, M. and {Korn}, A.~J. and {Lanzafame}, A.~C. and {Manteiga}, M. and {Moitinho}, A. and {Muinonen}, K. and {Osinde}, J. and {Pancino}, E. and {Pauwels}, T. and {Petit}, J. -M. and {Recio-Blanco}, A. and {Robin}, A.~C. and {Sarro}, L.~M. and {Siopis}, C. and {Smith}, M. and {Smith}, K.~W. and {Sozzetti}, A. and {Thuillot}, W. and {van Reeven}, W. and {Viala}, Y. and {Abbas}, U. and {Abreu Aramburu}, A. and {Accart}, S. and {Aguado}, J.~J. and {Allan}, P.~M. and {Allasia}, W. and {Altavilla}, G. and {{\'A}lvarez}, M.~A. and {Alves}, J. and {Anderson}, R.~I. and {Andrei}, A.~H. and {Anglada Varela}, E. and {Antiche}, E. and {Antoja}, T. and {Ant{\'o}n}, S. and {Arcay}, B. and {Atzei}, A. and {Ayache}, L. and {Bach}, N. and {Baker}, S.~G. and {Balaguer-N{\'u}{\~n}ez}, L. and {Barache}, C. and {Barata}, C. and {Barbier}, A. and {Barblan}, F. and {Baroni}, M. and {Barrado y Navascu{\'e}s}, D. and {Barros}, M. and {Barstow}, M.~A. and {Becciani}, U. and {Bellazzini}, M. and {Bellei}, G. and {Bello Garc{\'\i}a}, A. and {Belokurov}, V. and {Bendjoya}, P. and {Berihuete}, A. and {Bianchi}, L. and {Bienaym{\'e}}, O. and {Billebaud}, F. and {Blagorodnova}, N. and {Blanco-Cuaresma}, S. and {Boch}, T. and {Bombrun}, A. and {Borrachero}, R. and {Bouquillon}, S. and {Bourda}, G. and {Bouy}, H. and {Bragaglia}, A. and {Breddels}, M.~A. and {Brouillet}, N. and {Br{\"u}semeister}, T. and {Bucciarelli}, B. and {Budnik}, F. and {Burgess}, P. and {Burgon}, R. and {Burlacu}, A. and {Busonero}, D. and {Buzzi}, R. and {Caffau}, E. and {Cambras}, J. and {Campbell}, H. and {Cancelliere}, R. and {Cantat-Gaudin}, T. and {Carlucci}, T. and {Carrasco}, J.~M. and {Castellani}, M. and {Charlot}, P. and {Charnas}, J. and {Charvet}, P. and {Chassat}, F. and {Chiavassa}, A. and {Clotet}, M. and {Cocozza}, G. and {Collins}, R.~S. and {Collins}, P. and {Costigan}, G.},
        title = "{The Gaia mission}",
      journal = {\aap},
         year = 2016,
        month = nov,
       volume = {595},
          eid = {A1},
        pages = {A1},
          doi = {10.1051/0004-6361/201629272},
archivePrefix = {arXiv},
       eprint = {1609.04153},
 primaryClass = {astro-ph.IM},
       adsurl = {https://ui.adsabs.harvard.edu/abs/2016A&A...595A...1G}
}

@ARTICLE{2018A&A...616A...1G,
       author = {{Gaia Collaboration} and {Brown}, A.~G.~A. and {Vallenari}, A. and {Prusti}, T. and {de Bruijne}, J.~H.~J. and {Babusiaux}, C. and {Bailer-Jones}, C.~A.~L. and {Biermann}, M. and {Evans}, D.~W. and {Eyer}, L. and {Jansen}, F. and {Jordi}, C. and {Klioner}, S.~A. and {Lammers}, U. and {Lindegren}, L. and {Luri}, X. and {Mignard}, F. and {Panem}, C. and {Pourbaix}, D. and {Randich}, S. and {Sartoretti}, P. and {Siddiqui}, H.~I. and {Soubiran}, C. and {van Leeuwen}, F. and {Walton}, N.~A. and {Arenou}, F. and {Bastian}, U. and {Cropper}, M. and {Drimmel}, R. and {Katz}, D. and {Lattanzi}, M.~G. and {Bakker}, J. and {Cacciari}, C. and {Casta{\~n}eda}, J. and {Chaoul}, L. and {Cheek}, N. and {De Angeli}, F. and {Fabricius}, C. and {Guerra}, R. and {Holl}, B. and {Masana}, E. and {Messineo}, R. and {Mowlavi}, N. and {Nienartowicz}, K. and {Panuzzo}, P. and {Portell}, J. and {Riello}, M. and {Seabroke}, G.~M. and {Tanga}, P. and {Th{\'e}venin}, F. and {Gracia-Abril}, G. and {Comoretto}, G. and {Garcia-Reinaldos}, M. and {Teyssier}, D. and {Altmann}, M. and {Andrae}, R. and {Audard}, M. and {Bellas-Velidis}, I. and {Benson}, K. and {Berthier}, J. and {Blomme}, R. and {Burgess}, P. and {Busso}, G. and {Carry}, B. and {Cellino}, A. and {Clementini}, G. and {Clotet}, M. and {Creevey}, O. and {Davidson}, M. and {De Ridder}, J. and {Delchambre}, L. and {Dell'Oro}, A. and {Ducourant}, C. and {Fern{\'a}ndez-Hern{\'a}ndez}, J. and {Fouesneau}, M. and {Fr{\'e}mat}, Y. and {Galluccio}, L. and {Garc{\'\i}a-Torres}, M. and {Gonz{\'a}lez-N{\'u}{\~n}ez}, J. and {Gonz{\'a}lez-Vidal}, J.~J. and {Gosset}, E. and {Guy}, L.~P. and {Halbwachs}, J. -L. and {Hambly}, N.~C. and {Harrison}, D.~L. and {Hern{\'a}ndez}, J. and {Hestroffer}, D. and {Hodgkin}, S.~T. and {Hutton}, A. and {Jasniewicz}, G. and {Jean-Antoine-Piccolo}, A. and {Jordan}, S. and {Korn}, A.~J. and {Krone-Martins}, A. and {Lanzafame}, A.~C. and {Lebzelter}, T. and {L{\"o}ffler}, W. and {Manteiga}, M. and {Marrese}, P.~M. and {Mart{\'\i}n-Fleitas}, J.~M. and {Moitinho}, A. and {Mora}, A. and {Muinonen}, K. and {Osinde}, J. and {Pancino}, E. and {Pauwels}, T. and {Petit}, J. -M. and {Recio-Blanco}, A. and {Richards}, P.~J. and {Rimoldini}, L. and {Robin}, A.~C. and {Sarro}, L.~M. and {Siopis}, C. and {Smith}, M. and {Sozzetti}, A. and {S{\"u}veges}, M. and {Torra}, J. and {van Reeven}, W. and {Abbas}, U. and {Abreu Aramburu}, A. and {Accart}, S. and {Aerts}, C. and {Altavilla}, G. and {{\'A}lvarez}, M.~A. and {Alvarez}, R. and {Alves}, J. and {Anderson}, R.~I. and {Andrei}, A.~H. and {Anglada Varela}, E. and {Antiche}, E. and {Antoja}, T. and {Arcay}, B. and {Astraatmadja}, T.~L. and {Bach}, N. and {Baker}, S.~G. and {Balaguer-N{\'u}{\~n}ez}, L. and {Balm}, P. and {Barache}, C. and {Barata}, C. and {Barbato}, D. and {Barblan}, F. and {Barklem}, P.~S. and {Barrado}, D. and {Barros}, M. and {Barstow}, M.~A. and {Bartholom{\'e} Mu{\~n}oz}, S. and {Bassilana}, J. -L. and {Becciani}, U. and {Bellazzini}, M. and {Berihuete}, A. and {Bertone}, S. and {Bianchi}, L. and {Bienaym{\'e}}, O. and {Blanco-Cuaresma}, S. and {Boch}, T. and {Boeche}, C. and {Bombrun}, A. and {Borrachero}, R. and {Bossini}, D. and {Bouquillon}, S. and {Bourda}, G. and {Bragaglia}, A. and {Bramante}, L. and {Breddels}, M.~A. and {Bressan}, A. and {Brouillet}, N. and {Br{\"u}semeister}, T. and {Brugaletta}, E. and {Bucciarelli}, B. and {Burlacu}, A. and {Busonero}, D. and {Butkevich}, A.~G. and {Buzzi}, R. and {Caffau}, E. and {Cancelliere}, R. and {Cannizzaro}, G. and {Cantat-Gaudin}, T. and {Carballo}, R. and {Carlucci}, T. and {Carrasco}, J.~M. and {Casamiquela}, L. and {Castellani}, M. and {Castro-Ginard}, A. and {Charlot}, P. and {Chemin}, L. and {Chiavassa}, A. and {Cocozza}, G. and {Costigan}, G. and {Cowell}, S. and {Crifo}, F. and {Crosta}, M. and {Crowley}, C. and {Cuypers}, J. and {Dafonte}, C. and {Damerdji}, Y. and {Dapergolas}, A. and {David}, P. and {David}, M. and {de Laverny}, P. and {De Luise}, F.},
        title = "{Gaia Data Release 2. Summary of the contents and survey properties}",
      journal = {\aap},
         year = 2018,
        month = aug,
       volume = {616},
          eid = {A1},
        pages = {A1},
          doi = {10.1051/0004-6361/201833051},
archivePrefix = {arXiv},
       eprint = {1804.09365},
 primaryClass = {astro-ph.GA},
       adsurl = {https://ui.adsabs.harvard.edu/abs/2018A&A...616A...1G}
}

@ARTICLE{2021A&A...649A...1G,
       author = {{Gaia Collaboration} and {Brown}, A.~G.~A. and {Vallenari}, A. and {Prusti}, T. and {de Bruijne}, J.~H.~J. and {Babusiaux}, C. and {Biermann}, M. and {Creevey}, O.~L. and {Evans}, D.~W. and {Eyer}, L. and {Hutton}, A. and {Jansen}, F. and {Jordi}, C. and {Klioner}, S.~A. and {Lammers}, U. and {Lindegren}, L. and {Luri}, X. and {Mignard}, F. and {Panem}, C. and {Pourbaix}, D. and {Randich}, S. and {Sartoretti}, P. and {Soubiran}, C. and {Walton}, N.~A. and {Arenou}, F. and {Bailer-Jones}, C.~A.~L. and {Bastian}, U. and {Cropper}, M. and {Drimmel}, R. and {Katz}, D. and {Lattanzi}, M.~G. and {van Leeuwen}, F. and {Bakker}, J. and {Cacciari}, C. and {Casta{\~n}eda}, J. and {De Angeli}, F. and {Ducourant}, C. and {Fabricius}, C. and {Fouesneau}, M. and {Fr{\'e}mat}, Y. and {Guerra}, R. and {Guerrier}, A. and {Guiraud}, J. and {Jean-Antoine Piccolo}, A. and {Masana}, E. and {Messineo}, R. and {Mowlavi}, N. and {Nicolas}, C. and {Nienartowicz}, K. and {Pailler}, F. and {Panuzzo}, P. and {Riclet}, F. and {Roux}, W. and {Seabroke}, G.~M. and {Sordo}, R. and {Tanga}, P. and {Th{\'e}venin}, F. and {Gracia-Abril}, G. and {Portell}, J. and {Teyssier}, D. and {Altmann}, M. and {Andrae}, R. and {Bellas-Velidis}, I. and {Benson}, K. and {Berthier}, J. and {Blomme}, R. and {Brugaletta}, E. and {Burgess}, P.~W. and {Busso}, G. and {Carry}, B. and {Cellino}, A. and {Cheek}, N. and {Clementini}, G. and {Damerdji}, Y. and {Davidson}, M. and {Delchambre}, L. and {Dell'Oro}, A. and {Fern{\'a}ndez-Hern{\'a}ndez}, J. and {Galluccio}, L. and {Garc{\'\i}a-Lario}, P. and {Garcia-Reinaldos}, M. and {Gonz{\'a}lez-N{\'u}{\~n}ez}, J. and {Gosset}, E. and {Haigron}, R. and {Halbwachs}, J. -L. and {Hambly}, N.~C. and {Harrison}, D.~L. and {Hatzidimitriou}, D. and {Heiter}, U. and {Hern{\'a}ndez}, J. and {Hestroffer}, D. and {Hodgkin}, S.~T. and {Holl}, B. and {Jan{\ss}en}, K. and {Jevardat de Fombelle}, G. and {Jordan}, S. and {Krone-Martins}, A. and {Lanzafame}, A.~C. and {L{\"o}ffler}, W. and {Lorca}, A. and {Manteiga}, M. and {Marchal}, O. and {Marrese}, P.~M. and {Moitinho}, A. and {Mora}, A. and {Muinonen}, K. and {Osborne}, P. and {Pancino}, E. and {Pauwels}, T. and {Petit}, J. -M. and {Recio-Blanco}, A. and {Richards}, P.~J. and {Riello}, M. and {Rimoldini}, L. and {Robin}, A.~C. and {Roegiers}, T. and {Rybizki}, J. and {Sarro}, L.~M. and {Siopis}, C. and {Smith}, M. and {Sozzetti}, A. and {Ulla}, A. and {Utrilla}, E. and {van Leeuwen}, M. and {van Reeven}, W. and {Abbas}, U. and {Abreu Aramburu}, A. and {Accart}, S. and {Aerts}, C. and {Aguado}, J.~J. and {Ajaj}, M. and {Altavilla}, G. and {{\'A}lvarez}, M.~A. and {{\'A}lvarez Cid-Fuentes}, J. and {Alves}, J. and {Anderson}, R.~I. and {Anglada Varela}, E. and {Antoja}, T. and {Audard}, M. and {Baines}, D. and {Baker}, S.~G. and {Balaguer-N{\'u}{\~n}ez}, L. and {Balbinot}, E. and {Balog}, Z. and {Barache}, C. and {Barbato}, D. and {Barros}, M. and {Barstow}, M.~A. and {Bartolom{\'e}}, S. and {Bassilana}, J. -L. and {Bauchet}, N. and {Baudesson-Stella}, A. and {Becciani}, U. and {Bellazzini}, M. and {Bernet}, M. and {Bertone}, S. and {Bianchi}, L. and {Blanco-Cuaresma}, S. and {Boch}, T. and {Bombrun}, A. and {Bossini}, D. and {Bouquillon}, S. and {Bragaglia}, A. and {Bramante}, L. and {Breedt}, E. and {Bressan}, A. and {Brouillet}, N. and {Bucciarelli}, B. and {Burlacu}, A. and {Busonero}, D. and {Butkevich}, A.~G. and {Buzzi}, R. and {Caffau}, E. and {Cancelliere}, R. and {C{\'a}novas}, H. and {Cantat-Gaudin}, T. and {Carballo}, R. and {Carlucci}, T. and {Carnerero}, M.~I. and {Carrasco}, J.~M. and {Casamiquela}, L. and {Castellani}, M. and {Castro-Ginard}, A. and {Castro Sampol}, P. and {Chaoul}, L. and {Charlot}, P. and {Chemin}, L. and {Chiavassa}, A. and {Cioni}, M. -R.~L. and {Comoretto}, G. and {Cooper}, W.~J. and {Cornez}, T. and {Cowell}, S. and {Crifo}, F. and {Crosta}, M. and {Crowley}, C. and {Dafonte}, C. and {Dapergolas}, A. and {David}, M. and {David}, P.},
        title = "{Gaia Early Data Release 3. Summary of the contents and survey properties}",
      journal = {\aap},
         year = 2021,
        month = may,
       volume = {649},
          eid = {A1},
        pages = {A1},
          doi = {10.1051/0004-6361/202039657},
archivePrefix = {arXiv},
       eprint = {2012.01533},
 primaryClass = {astro-ph.GA},
       adsurl = {https://ui.adsabs.harvard.edu/abs/2021A&A...649A...1G}
}

@ARTICLE{2023A&A...674A...1G,
       author = {{Gaia Collaboration} and {Vallenari}, A. and {Brown}, A.~G.~A. and {Prusti}, T. and {de Bruijne}, J.~H.~J. and {Arenou}, F. and {Babusiaux}, C. and {Biermann}, M. and {Creevey}, O.~L. and {Ducourant}, C. and {Evans}, D.~W. and {Eyer}, L. and {Guerra}, R. and {Hutton}, A. and {Jordi}, C. and {Klioner}, S.~A. and {Lammers}, U.~L. and {Lindegren}, L. and {Luri}, X. and {Mignard}, F. and {Panem}, C. and {Pourbaix}, D. and {Randich}, S. and {Sartoretti}, P. and {Soubiran}, C. and {Tanga}, P. and {Walton}, N.~A. and {Bailer-Jones}, C.~A.~L. and {Bastian}, U. and {Drimmel}, R. and {Jansen}, F. and {Katz}, D. and {Lattanzi}, M.~G. and {van Leeuwen}, F. and {Bakker}, J. and {Cacciari}, C. and {Casta{\~n}eda}, J. and {De Angeli}, F. and {Fabricius}, C. and {Fouesneau}, M. and {Fr{\'e}mat}, Y. and {Galluccio}, L. and {Guerrier}, A. and {Heiter}, U. and {Masana}, E. and {Messineo}, R. and {Mowlavi}, N. and {Nicolas}, C. and {Nienartowicz}, K. and {Pailler}, F. and {Panuzzo}, P. and {Riclet}, F. and {Roux}, W. and {Seabroke}, G.~M. and {Sordo}, R. and {Th{\'e}venin}, F. and {Gracia-Abril}, G. and {Portell}, J. and {Teyssier}, D. and {Altmann}, M. and {Andrae}, R. and {Audard}, M. and {Bellas-Velidis}, I. and {Benson}, K. and {Berthier}, J. and {Blomme}, R. and {Burgess}, P.~W. and {Busonero}, D. and {Busso}, G. and {C{\'a}novas}, H. and {Carry}, B. and {Cellino}, A. and {Cheek}, N. and {Clementini}, G. and {Damerdji}, Y. and {Davidson}, M. and {de Teodoro}, P. and {Nu{\~n}ez Campos}, M. and {Delchambre}, L. and {Dell'Oro}, A. and {Esquej}, P. and {Fern{\'a}ndez-Hern{\'a}ndez}, J. and {Fraile}, E. and {Garabato}, D. and {Garc{\'\i}a-Lario}, P. and {Gosset}, E. and {Haigron}, R. and {Halbwachs}, J. -L. and {Hambly}, N.~C. and {Harrison}, D.~L. and {Hern{\'a}ndez}, J. and {Hestroffer}, D. and {Hodgkin}, S.~T. and {Holl}, B. and {Jan{\ss}en}, K. and {Jevardat de Fombelle}, G. and {Jordan}, S. and {Krone-Martins}, A. and {Lanzafame}, A.~C. and {L{\"o}ffler}, W. and {Marchal}, O. and {Marrese}, P.~M. and {Moitinho}, A. and {Muinonen}, K. and {Osborne}, P. and {Pancino}, E. and {Pauwels}, T. and {Recio-Blanco}, A. and {Reyl{\'e}}, C. and {Riello}, M. and {Rimoldini}, L. and {Roegiers}, T. and {Rybizki}, J. and {Sarro}, L.~M. and {Siopis}, C. and {Smith}, M. and {Sozzetti}, A. and {Utrilla}, E. and {van Leeuwen}, M. and {Abbas}, U. and {{\'A}brah{\'a}m}, P. and {Abreu Aramburu}, A. and {Aerts}, C. and {Aguado}, J.~J. and {Ajaj}, M. and {Aldea-Montero}, F. and {Altavilla}, G. and {{\'A}lvarez}, M.~A. and {Alves}, J. and {Anders}, F. and {Anderson}, R.~I. and {Anglada Varela}, E. and {Antoja}, T. and {Baines}, D. and {Baker}, S.~G. and {Balaguer-N{\'u}{\~n}ez}, L. and {Balbinot}, E. and {Balog}, Z. and {Barache}, C. and {Barbato}, D. and {Barros}, M. and {Barstow}, M.~A. and {Bartolom{\'e}}, S. and {Bassilana}, J. -L. and {Bauchet}, N. and {Becciani}, U. and {Bellazzini}, M. and {Berihuete}, A. and {Bernet}, M. and {Bertone}, S. and {Bianchi}, L. and {Binnenfeld}, A. and {Blanco-Cuaresma}, S. and {Blazere}, A. and {Boch}, T. and {Bombrun}, A. and {Bossini}, D. and {Bouquillon}, S. and {Bragaglia}, A. and {Bramante}, L. and {Breedt}, E. and {Bressan}, A. and {Brouillet}, N. and {Brugaletta}, E. and {Bucciarelli}, B. and {Burlacu}, A. and {Butkevich}, A.~G. and {Buzzi}, R. and {Caffau}, E. and {Cancelliere}, R. and {Cantat-Gaudin}, T. and {Carballo}, R. and {Carlucci}, T. and {Carnerero}, M.~I. and {Carrasco}, J.~M. and {Casamiquela}, L. and {Castellani}, M. and {Castro-Ginard}, A. and {Chaoul}, L. and {Charlot}, P. and {Chemin}, L. and {Chiaramida}, V. and {Chiavassa}, A. and {Chornay}, N. and {Comoretto}, G. and {Contursi}, G. and {Cooper}, W.~J. and {Cornez}, T. and {Cowell}, S. and {Crifo}, F. and {Cropper}, M. and {Crosta}, M. and {Crowley}, C. and {Dafonte}, C. and {Dapergolas}, A. and {David}, M. and {David}, P. and {de Laverny}, P. and {De Luise}, F. and {De March}, R.},
        title = "{Gaia Data Release 3. Summary of the content and survey properties}",
      journal = {\aap},
         year = 2023,
        month = jun,
       volume = {674},
          eid = {A1},
        pages = {A1},
          doi = {10.1051/0004-6361/202243940},
archivePrefix = {arXiv},
       eprint = {2208.00211},
 primaryClass = {astro-ph.GA},
       adsurl = {https://ui.adsabs.harvard.edu/abs/2023A&A...674A...1G}
}

@ARTICLE{2018AJ....156..123A,
       author = {{Astropy Collaboration} and {Price-Whelan}, A.~M. and {Sip{\H{o}}cz}, B.~M. and {G{\"u}nther}, H.~M. and {Lim}, P.~L. and {Crawford}, S.~M. and {Conseil}, S. and {Shupe}, D.~L. and {Craig}, M.~W. and {Dencheva}, N. and {Ginsburg}, A. and {VanderPlas}, J.~T. and {Bradley}, L.~D. and {P{\'e}rez-Su{\'a}rez}, D. and {de Val-Borro}, M. and {Aldcroft}, T.~L. and {Cruz}, K.~L. and {Robitaille}, T.~P. and {Tollerud}, E.~J. and {Ardelean}, C. and {Babej}, T. and {Bach}, Y.~P. and {Bachetti}, M. and {Bakanov}, A.~V. and {Bamford}, S.~P. and {Barentsen}, G. and {Barmby}, P. and {Baumbach}, A. and {Berry}, K.~L. and {Biscani}, F. and {Boquien}, M. and {Bostroem}, K.~A. and {Bouma}, L.~G. and {Brammer}, G.~B. and {Bray}, E.~M. and {Breytenbach}, H. and {Buddelmeijer}, H. and {Burke}, D.~J. and {Calderone}, G. and {Cano Rodr{\'\i}guez}, J.~L. and {Cara}, M. and {Cardoso}, J.~V.~M. and {Cheedella}, S. and {Copin}, Y. and {Corrales}, L. and {Crichton}, D. and {D'Avella}, D. and {Deil}, C. and {Depagne}, {\'E}. and {Dietrich}, J.~P. and {Donath}, A. and {Droettboom}, M. and {Earl}, N. and {Erben}, T. and {Fabbro}, S. and {Ferreira}, L.~A. and {Finethy}, T. and {Fox}, R.~T. and {Garrison}, L.~H. and {Gibbons}, S.~L.~J. and {Goldstein}, D.~A. and {Gommers}, R. and {Greco}, J.~P. and {Greenfield}, P. and {Groener}, A.~M. and {Grollier}, F. and {Hagen}, A. and {Hirst}, P. and {Homeier}, D. and {Horton}, A.~J. and {Hosseinzadeh}, G. and {Hu}, L. and {Hunkeler}, J.~S. and {Ivezi{\'c}}, {\v{Z}}. and {Jain}, A. and {Jenness}, T. and {Kanarek}, G. and {Kendrew}, S. and {Kern}, N.~S. and {Kerzendorf}, W.~E. and {Khvalko}, A. and {King}, J. and {Kirkby}, D. and {Kulkarni}, A.~M. and {Kumar}, A. and {Lee}, A. and {Lenz}, D. and {Littlefair}, S.~P. and {Ma}, Z. and {Macleod}, D.~M. and {Mastropietro}, M. and {McCully}, C. and {Montagnac}, S. and {Morris}, B.~M. and {Mueller}, M. and {Mumford}, S.~J. and {Muna}, D. and {Murphy}, N.~A. and {Nelson}, S. and {Nguyen}, G.~H. and {Ninan}, J.~P. and {N{\"o}the}, M. and {Ogaz}, S. and {Oh}, S. and {Parejko}, J.~K. and {Parley}, N. and {Pascual}, S. and {Patil}, R. and {Patil}, A.~A. and {Plunkett}, A.~L. and {Prochaska}, J.~X. and {Rastogi}, T. and {Reddy Janga}, V. and {Sabater}, J. and {Sakurikar}, P. and {Seifert}, M. and {Sherbert}, L.~E. and {Sherwood-Taylor}, H. and {Shih}, A.~Y. and {Sick}, J. and {Silbiger}, M.~T. and {Singanamalla}, S. and {Singer}, L.~P. and {Sladen}, P.~H. and {Sooley}, K.~A. and {Sornarajah}, S. and {Streicher}, O. and {Teuben}, P. and {Thomas}, S.~W. and {Tremblay}, G.~R. and {Turner}, J.~E.~H. and {Terr{\'o}n}, V. and {van Kerkwijk}, M.~H. and {de la Vega}, A. and {Watkins}, L.~L. and {Weaver}, B.~A. and {Whitmore}, J.~B. and {Woillez}, J. and {Zabalza}, V. and {Astropy Contributors}},
        title = "{The Astropy Project: Building an Open-science Project and Status of the v2.0 Core Package}",
      journal = {\aj},
         year = 2018,
        month = sep,
       volume = {156},
       number = {3},
          eid = {123},
        pages = {123},
          doi = {10.3847/1538-3881/aabc4f},
archivePrefix = {arXiv},
       eprint = {1801.02634},
 primaryClass = {astro-ph.IM},
       adsurl = {https://ui.adsabs.harvard.edu/abs/2018AJ....156..123A}
}

@ARTICLE{2013A&A...558A..33A,
       author = {{Astropy Collaboration} and {Robitaille}, Thomas P. and {Tollerud}, Erik J. and {Greenfield}, Perry and {Droettboom}, Michael and {Bray}, Erik and {Aldcroft}, Tom and {Davis}, Matt and {Ginsburg}, Adam and {Price-Whelan}, Adrian M. and {Kerzendorf}, Wolfgang E. and {Conley}, Alexander and {Crighton}, Neil and {Barbary}, Kyle and {Muna}, Demitri and {Ferguson}, Henry and {Grollier}, Fr{\'e}d{\'e}ric and {Parikh}, Madhura M. and {Nair}, Prasanth H. and {Unther}, Hans M. and {Deil}, Christoph and {Woillez}, Julien and {Conseil}, Simon and {Kramer}, Roban and {Turner}, James E.~H. and {Singer}, Leo and {Fox}, Ryan and {Weaver}, Benjamin A. and {Zabalza}, Victor and {Edwards}, Zachary I. and {Azalee Bostroem}, K. and {Burke}, D.~J. and {Casey}, Andrew R. and {Crawford}, Steven M. and {Dencheva}, Nadia and {Ely}, Justin and {Jenness}, Tim and {Labrie}, Kathleen and {Lim}, Pey Lian and {Pierfederici}, Francesco and {Pontzen}, Andrew and {Ptak}, Andy and {Refsdal}, Brian and {Servillat}, Mathieu and {Streicher}, Ole},
        title = "{Astropy: A community Python package for astronomy}",
      journal = {\aap},
         year = 2013,
        month = oct,
       volume = {558},
          eid = {A33},
        pages = {A33},
          doi = {10.1051/0004-6361/201322068},
archivePrefix = {arXiv},
       eprint = {1307.6212},
 primaryClass = {astro-ph.IM},
       adsurl = {https://ui.adsabs.harvard.edu/abs/2013A&A...558A..33A}
}

@ARTICLE{2023A&A...673A.114H,
       author = {{Hunt}, Emily L. and {Reffert}, Sabine},
        title = "{Improving the open cluster census. II. An all-sky cluster catalogue with Gaia DR3}",
      journal = {\aap},
         year = 2023,
        month = may,
       volume = {673},
          eid = {A114},
        pages = {A114},
          doi = {10.1051/0004-6361/202346285},
archivePrefix = {arXiv},
       eprint = {2303.13424},
 primaryClass = {astro-ph.GA},
       adsurl = {https://ui.adsabs.harvard.edu/abs/2023A&A...673A.114H}
}

@ARTICLE{2021MNRAS.508.3877G,
       author = {{Gentile Fusillo}, N.~P. and {Tremblay}, P.-E. and {Cukanovaite}, E. and {Vorontseva}, A. and {Lallement}, R. and {Hollands}, M. and {G{\"a}nsicke}, B.~T. and {Burdge}, K.~B. and {McCleery}, J. and {Jordan}, S.},
        title = "{A catalogue of white dwarfs in Gaia EDR3}",
      journal = {\mnras},
         year = 2021,
        month = dec,
       volume = {508},
       number = {3},
        pages = {3877-3896},
          doi = {10.1093/mnras/stab2672},
archivePrefix = {arXiv},
       eprint = {2106.07669},
 primaryClass = {astro-ph.SR},
       adsurl = {https://ui.adsabs.harvard.edu/abs/2021MNRAS.508.3877G}
}

@ARTICLE{2012MNRAS.427..127B,
       author = {{Bressan}, Alessandro and {Marigo}, Paola and {Girardi}, L{\'e}o. and {Salasnich}, Bernardo and {Dal Cero}, Claudia and {Rubele}, Stefano and {Nanni}, Ambra},
        title = "{PARSEC: stellar tracks and isochrones with the PAdova and TRieste Stellar Evolution Code}",
      journal = {\mnras},
         year = 2012,
        month = nov,
       volume = {427},
       number = {1},
        pages = {127-145},
          doi = {10.1111/j.1365-2966.2012.21948.x},
archivePrefix = {arXiv},
       eprint = {1208.4498},
 primaryClass = {astro-ph.SR},
       adsurl = {https://ui.adsabs.harvard.edu/abs/2012MNRAS.427..127B}
}

@ARTICLE{2014MNRAS.444.2525C,
       author = {{Chen}, Yang and {Girardi}, L{\'e}o and {Bressan}, Alessandro and {Marigo}, Paola and {Barbieri}, Mauro and {Kong}, Xu},
        title = "{Improving PARSEC models for very low mass stars}",
      journal = {\mnras},
         year = 2014,
        month = nov,
       volume = {444},
       number = {3},
        pages = {2525-2543},
          doi = {10.1093/mnras/stu1605},
archivePrefix = {arXiv},
       eprint = {1409.0322},
 primaryClass = {astro-ph.SR},
       adsurl = {https://ui.adsabs.harvard.edu/abs/2014MNRAS.444.2525C}
}

@ARTICLE{2015MNRAS.454.2787G,
       author = {{Guo}, Jincheng and {Zhao}, Jingkun and {Tziamtzis}, Anestis and {Liu}, Jifeng and {Li}, Lifang and {Zhang}, Yong and {Hou}, Yonghui and {Wang}, Yuefei},
        title = "{White dwarfs identified in LAMOST DR 2}",
      journal = {\mnras},
         year = 2015,
        month = dec,
       volume = {454},
       number = {3},
        pages = {2787-2797},
          doi = {10.1093/mnras/stv2104},
       adsurl = {https://ui.adsabs.harvard.edu/abs/2015MNRAS.454.2787G}
}

@ARTICLE{2013ApJS..204....5K,
       author = {{Kleinman}, S.~J. and {Kepler}, S.~O. and {Koester}, D. and {Pelisoli}, Ingrid and {Pe{\c{c}}anha}, Viviane and {Nitta}, A. and {Costa}, J.~E.~S. and {Krzesinski}, J. and {Dufour}, P. and {Lachapelle}, F. -R. and {Bergeron}, P. and {Yip}, Ching-Wa and {Harris}, Hugh C. and {Eisenstein}, Daniel J. and {Althaus}, L. and {C{\'o}rsico}, A.},
        title = "{SDSS DR7 White Dwarf Catalog}",
      journal = {\apjs},
         year = 2013,
        month = jan,
       volume = {204},
       number = {1},
          eid = {5},
        pages = {5},
          doi = {10.1088/0067-0049/204/1/5},
archivePrefix = {arXiv},
       eprint = {1212.1222},
 primaryClass = {astro-ph.SR},
       adsurl = {https://ui.adsabs.harvard.edu/abs/2013ApJS..204....5K}
}

@ARTICLE{2018ApJ...866...21C,
       author = {{Cummings}, Jeffrey D. and {Kalirai}, Jason S. and {Tremblay}, P. -E. and {Ramirez-Ruiz}, Enrico and {Choi}, Jieun},
        title = "{The White Dwarf Initial-Final Mass Relation for Progenitor Stars from 0.85 to 7.5 M $_{{\ensuremath{\odot}}}$}",
      journal = {\apj},
         year = 2018,
        month = oct,
       volume = {866},
       number = {1},
          eid = {21},
        pages = {21},
          doi = {10.3847/1538-4357/aadfd6},
archivePrefix = {arXiv},
       eprint = {1809.01673},
 primaryClass = {astro-ph.SR},
       adsurl = {https://ui.adsabs.harvard.edu/abs/2018ApJ...866...21C}
}

@ARTICLE{2020ApJ...901...93B,
       author = {{B{\'e}dard}, A. and {Bergeron}, P. and {Brassard}, P. and {Fontaine}, G.},
        title = "{On the Spectral Evolution of Hot White Dwarf Stars. I. A Detailed Model Atmosphere Analysis of Hot White Dwarfs from SDSS DR12}",
      journal = {\apj},
         year = 2020,
        month = oct,
       volume = {901},
       number = {2},
          eid = {93},
        pages = {93},
          doi = {10.3847/1538-4357/abafbe},
archivePrefix = {arXiv},
       eprint = {2008.07469},
 primaryClass = {astro-ph.SR},
       adsurl = {https://ui.adsabs.harvard.edu/abs/2020ApJ...901...93B}
}

@ARTICLE{2021A&A...645A..13P,
       author = {{Pri{\v{s}}egen}, M. and {Piecka}, M. and {Faltov{\'a}}, N. and {Kajan}, M. and {Paunzen}, E.},
        title = "{White dwarf-open cluster associations based on Gaia DR2}",
      journal = {\aap},
         year = 2021,
        month = jan,
       volume = {645},
          eid = {A13},
        pages = {A13},
          doi = {10.1051/0004-6361/202039276},
       adsurl = {https://ui.adsabs.harvard.edu/abs/2021A&A...645A..13P}
}

@ARTICLE{2019MNRAS.486.2169K,
       author = {{Kepler}, S.~O. and {Pelisoli}, Ingrid and {Koester}, Detlev and {Reindl}, Nicole and {Geier}, Stephan and {Romero}, Alejandra D. and {Ourique}, Gustavo and {Oliveira}, Cristiane de Paula and {Amaral}, Larissa A.},
        title = "{White dwarf and subdwarf stars in the Sloan Digital Sky Survey Data Release 14}",
      journal = {\mnras},
         year = 2019,
        month = jun,
       volume = {486},
       number = {2},
        pages = {2169-2183},
          doi = {10.1093/mnras/stz960},
archivePrefix = {arXiv},
       eprint = {1904.01626},
 primaryClass = {astro-ph.SR},
       adsurl = {https://ui.adsabs.harvard.edu/abs/2019MNRAS.486.2169K}
}

@ARTICLE{2021ApJ...912..165R,
       author = {{Richer}, Harvey B. and {Caiazzo}, Ilaria and {Du}, Helen and {Grondin}, Steffani and {Hegarty}, James and {Heyl}, Jeremy and {Kerr}, Ronan and {Miller}, David R. and {Thiele}, Sarah},
        title = "{Massive White Dwarfs in Young Star Clusters}",
      journal = {\apj},
         year = 2021,
        month = may,
       volume = {912},
       number = {2},
          eid = {165},
        pages = {165},
          doi = {10.3847/1538-4357/abdeb7},
archivePrefix = {arXiv},
       eprint = {2101.08300},
 primaryClass = {astro-ph.SR},
       adsurl = {https://ui.adsabs.harvard.edu/abs/2021ApJ...912..165R}
}

@ARTICLE{2022ApJ...926..132H,
       author = {{Heyl}, Jeremy and {Caiazzo}, Ilaria and {Richer}, Harvey B.},
        title = "{Reconstructing the Pleiades with Gaia EDR3}",
      journal = {\apj},
         year = 2022,
        month = feb,
       volume = {926},
       number = {2},
          eid = {132},
        pages = {132},
          doi = {10.3847/1538-4357/ac45fc},
archivePrefix = {arXiv},
       eprint = {2110.03837},
 primaryClass = {astro-ph.SR},
       adsurl = {https://ui.adsabs.harvard.edu/abs/2022ApJ...926..132H}
}

@ARTICLE{2022ApJ...926L..24M,
       author = {{Miller}, David R. and {Caiazzo}, Ilaria and {Heyl}, Jeremy and {Richer}, Harvey B. and {Tremblay}, Pier-Emmanuel},
        title = "{The Ultramassive White Dwarfs of the Alpha Persei Cluster}",
      journal = {\apjl},
         year = 2022,
        month = feb,
       volume = {926},
       number = {2},
          eid = {L24},
        pages = {L24},
          doi = {10.3847/2041-8213/ac50a5},
archivePrefix = {arXiv},
       eprint = {2110.09668},
 primaryClass = {astro-ph.SR},
       adsurl = {https://ui.adsabs.harvard.edu/abs/2022ApJ...926L..24M}
}

@ARTICLE{2020ApJ...901L..14C,
       author = {{Caiazzo}, Ilaria and {Heyl}, Jeremy and {Richer}, Harvey and {Cummings}, Jeffrey and {Fleury}, Leesa and {Hegarty}, James and {Kalirai}, Jason and {Kerr}, Ronan and {Thiele}, Sarah and {Tremblay}, Pier-Emmanuel and {Villanueva}, Michael},
        title = "{Intermediate-mass Stars Become Magnetic White Dwarfs}",
      journal = {\apjl},
         year = 2020,
        month = sep,
       volume = {901},
       number = {1},
          eid = {L14},
        pages = {L14},
          doi = {10.3847/2041-8213/abb5f7},
archivePrefix = {arXiv},
       eprint = {2009.03374},
 primaryClass = {astro-ph.SR},
       adsurl = {https://ui.adsabs.harvard.edu/abs/2020ApJ...901L..14C}
}

@ARTICLE{2018MNRAS.480.4884E,
       author = {{El-Badry}, Kareem and {Rix}, Hans-Walter},
        title = "{Imprints of white dwarf recoil in the separation distribution of Gaia wide binaries}",
      journal = {\mnras},
         year = 2018,
        month = nov,
       volume = {480},
       number = {4},
        pages = {4884-4902},
          doi = {10.1093/mnras/sty2186},
archivePrefix = {arXiv},
       eprint = {1807.06011},
 primaryClass = {astro-ph.SR},
       adsurl = {https://ui.adsabs.harvard.edu/abs/2018MNRAS.480.4884E}
}

@ARTICLE{2020A&A...640A...1C,
       author = {{Cantat-Gaudin}, T. and {Anders}, F. and {Castro-Ginard}, A. and {Jordi}, C. and {Romero-G{\'o}mez}, M. and {Soubiran}, C. and {Casamiquela}, L. and {Tarricq}, Y. and {Moitinho}, A. and {Vallenari}, A. and {Bragaglia}, A. and {Krone-Martins}, A. and {Kounkel}, M.},
        title = "{Painting a portrait of the Galactic disc with its stellar clusters}",
      journal = {\aap},
         year = 2020,
        month = aug,
       volume = {640},
          eid = {A1},
        pages = {A1},
          doi = {10.1051/0004-6361/202038192},
archivePrefix = {arXiv},
       eprint = {2004.07274},
 primaryClass = {astro-ph.GA},
       adsurl = {https://ui.adsabs.harvard.edu/abs/2020A&A...640A...1C}
}

@ARTICLE{2022AJ....164..215B,
       author = {{Bouma}, L.~G. and {Kerr}, R. and {Curtis}, J.~L. and {Isaacson}, H. and {Hillenbrand}, L.~A. and {Howard}, A.~W. and {Kraus}, A.~L. and {Bieryla}, A. and {Latham}, D.~W. and {Petigura}, E.~A. and {Huber}, D.},
        title = "{Kepler and the Behemoth: Three Mini-Neptunes in a 40 Million Year Old Association}",
      journal = {\aj},
         year = 2022,
        month = nov,
       volume = {164},
       number = {5},
          eid = {215},
        pages = {215},
          doi = {10.3847/1538-3881/ac93ff},
archivePrefix = {arXiv},
       eprint = {2205.01112},
 primaryClass = {astro-ph.EP},
       adsurl = {https://ui.adsabs.harvard.edu/abs/2022AJ....164..215B}
}

@ARTICLE{2022A&A...665A.126N,
       author = {{Nguyen}, C.~T. and {Costa}, G. and {Girardi}, L. and {Volpato}, G. and {Bressan}, A. and {Chen}, Y. and {Marigo}, P. and {Fu}, X. and {Goudfrooij}, P.},
        title = "{PARSEC V2.0: Stellar tracks and isochrones of low- and intermediate-mass stars with rotation}",
      journal = {\aap},
         year = 2022,
        month = sep,
       volume = {665},
          eid = {A126},
        pages = {A126},
          doi = {10.1051/0004-6361/202244166},
archivePrefix = {arXiv},
       eprint = {2207.08642},
 primaryClass = {astro-ph.SR},
       adsurl = {https://ui.adsabs.harvard.edu/abs/2022A&A...665A.126N}
}

@ARTICLE{2011ApJ...730..128T,
       author = {{Tremblay}, P. -E. and {Bergeron}, P. and {Gianninas}, A.},
        title = "{An Improved Spectroscopic Analysis of DA White Dwarfs from the Sloan Digital Sky Survey Data Release 4}",
      journal = {\apj},
         year = 2011,
        month = apr,
       volume = {730},
       number = {2},
          eid = {128},
        pages = {128},
          doi = {10.1088/0004-637X/730/2/128},
archivePrefix = {arXiv},
       eprint = {1102.0056},
 primaryClass = {astro-ph.SR},
       adsurl = {https://ui.adsabs.harvard.edu/abs/2011ApJ...730..128T}
}

@ARTICLE{2007MNRAS.382..915H,
       author = {{Heyl}, Jeremy},
        title = "{Orbital evolution with white-dwarf kicks}",
      journal = {\mnras},
         year = 2007,
        month = dec,
       volume = {382},
       number = {2},
        pages = {915-920},
          doi = {10.1111/j.1365-2966.2007.12441.x},
archivePrefix = {arXiv},
       eprint = {0706.0912},
 primaryClass = {astro-ph},
       adsurl = {https://ui.adsabs.harvard.edu/abs/2007MNRAS.382..915H}
}

@ARTICLE{2003ApJ...595L..53F,
       author = {{Fellhauer}, M. and {Lin}, D.~N.~C. and {Bolte}, M. and {Aarseth}, S.~J. and {Williams}, K.~A.},
        title = "{The White Dwarf Deficit in Open Clusters: Dynamical Processes}",
      journal = {\apjl},
         year = 2003,
        month = sep,
       volume = {595},
       number = {1},
        pages = {L53-L56},
          doi = {10.1086/379005},
archivePrefix = {arXiv},
       eprint = {astro-ph/0308261},
 primaryClass = {astro-ph},
       adsurl = {https://ui.adsabs.harvard.edu/abs/2003ApJ...595L..53F}
}

@ARTICLE{2009ApJ...695L..20F,
       author = {{Fregeau}, John M. and {Richer}, Harvey B. and {Rasio}, Frederic A. and {Hurley}, Jarrod R.},
        title = "{The Dynamical Effects of White Dwarf Birth Kicks in Globular Star Clusters}",
      journal = {\apjl},
         year = 2009,
        month = apr,
       volume = {695},
       number = {1},
        pages = {L20-L24},
          doi = {10.1088/0004-637X/695/1/L20},
archivePrefix = {arXiv},
       eprint = {0902.1166},
 primaryClass = {astro-ph.GA},
       adsurl = {https://ui.adsabs.harvard.edu/abs/2009ApJ...695L..20F}
}

@ARTICLE{2018ApJ...861L..13G,
       author = {{Gagn{\'e}}, Jonathan and {Fontaine}, Gilles and {Simon}, Am{\'e}lie and {Faherty}, Jacqueline K.},
        title = "{A Young Ultramassive White Dwarf in the AB Doradus Moving Group}",
      journal = {\apjl},
         year = 2018,
        month = jul,
       volume = {861},
       number = {2},
          eid = {L13},
        pages = {L13},
          doi = {10.3847/2041-8213/aacdff},
archivePrefix = {arXiv},
       eprint = {1806.07512},
 primaryClass = {astro-ph.SR},
       adsurl = {https://ui.adsabs.harvard.edu/abs/2018ApJ...861L..13G}
}

@ARTICLE{2008AJ....135.2163R,
       author = {{Rubin}, Kate H.~R. and {Williams}, Kurtis A. and {Bolte}, M. and {Koester}, Detlev},
        title = "{The White Dwarf Population in NGC 1039 (M34) and the White Dwarf Initial-Final Mass Relation}",
      journal = {\aj},
         year = 2008,
        month = jun,
       volume = {135},
       number = {6},
        pages = {2163-2176},
          doi = {10.1088/0004-6256/135/6/2163},
archivePrefix = {arXiv},
       eprint = {0805.3156},
 primaryClass = {astro-ph},
       adsurl = {https://ui.adsabs.harvard.edu/abs/2008AJ....135.2163R}
}

@ARTICLE{2005ApJS..156...47L,
       author = {{Liebert}, James and {Bergeron}, P. and {Holberg}, J.~B.},
        title = "{The Formation Rate and Mass and Luminosity Functions of DA White Dwarfs from the Palomar Green Survey}",
      journal = {\apjs},
         year = 2005,
        month = jan,
       volume = {156},
       number = {1},
        pages = {47-68},
          doi = {10.1086/425738},
archivePrefix = {arXiv},
       eprint = {astro-ph/0406657},
 primaryClass = {astro-ph},
       adsurl = {https://ui.adsabs.harvard.edu/abs/2005ApJS..156...47L}
}

@ARTICLE{2005ApJ...618L.129K,
       author = {{Kalirai}, Jasonjot Singh and {Richer}, Harvey B. and {Hansen}, Brad M.~S. and {Reitzel}, David and {Rich}, R. Michael},
        title = "{The Dearth of Massive, Helium-rich White Dwarfs in Young Open Star Clusters}",
      journal = {\apjl},
         year = 2005,
        month = jan,
       volume = {618},
       number = {2},
        pages = {L129-L132},
          doi = {10.1086/427551},
       adsurl = {https://ui.adsabs.harvard.edu/abs/2005ApJ...618L.129K}
}

@ARTICLE{2021MNRAS.504..356D,
       author = {{Dias}, W.~S. and {Monteiro}, H. and {Moitinho}, A. and {L{\'e}pine}, J.~R.~D. and {Carraro}, G. and {Paunzen}, E. and {Alessi}, B. and {Villela}, L.},
        title = "{Updated parameters of 1743 open clusters based on Gaia DR2}",
      journal = {\mnras},
         year = 2021,
        month = jun,
       volume = {504},
       number = {1},
        pages = {356-371},
          doi = {10.1093/mnras/stab770},
archivePrefix = {arXiv},
       eprint = {2103.12829},
 primaryClass = {astro-ph.SR},
       adsurl = {https://ui.adsabs.harvard.edu/abs/2021MNRAS.504..356D}
}

@ARTICLE{2024A&A...686A..42H,
       author = {{Hunt}, Emily L. and {Reffert}, Sabine},
        title = "{Improving the open cluster census. III. Using cluster masses, radii, and dynamics to create a cleaned open cluster catalogue}",
      journal = {\aap},
         year = 2024,
        month = jun,
       volume = {686},
          eid = {A42},
        pages = {A42},
          doi = {10.1051/0004-6361/202348662},
archivePrefix = {arXiv},
       eprint = {2403.05143},
 primaryClass = {astro-ph.GA},
       adsurl = {https://ui.adsabs.harvard.edu/abs/2024A&A...686A..42H}
}

@ARTICLE{2022ApJ...935..167A,
       author = {{Astropy Collaboration} and {Price-Whelan}, Adrian M. and {Lim}, Pey Lian and {Earl}, Nicholas and {Starkman}, Nathaniel and {Bradley}, Larry and {Shupe}, David L. and {Patil}, Aarya A. and {Corrales}, Lia and {Brasseur}, C.~E. and {N{\"o}the}, Maximilian and {Donath}, Axel and {Tollerud}, Erik and {Morris}, Brett M. and {Ginsburg}, Adam and {Vaher}, Eero and {Weaver}, Benjamin A. and {Tocknell}, James and {Jamieson}, William and {van Kerkwijk}, Marten H. and {Robitaille}, Thomas P. and {Merry}, Bruce and {Bachetti}, Matteo and {G{\"u}nther}, H. Moritz and {Aldcroft}, Thomas L. and {Alvarado-Montes}, Jaime A. and {Archibald}, Anne M. and {B{\'o}di}, Attila and {Bapat}, Shreyas and {Barentsen}, Geert and {Baz{\'a}n}, Juanjo and {Biswas}, Manish and {Boquien}, M{\'e}d{\'e}ric and {Burke}, D.~J. and {Cara}, Daria and {Cara}, Mihai and {Conroy}, Kyle E. and {Conseil}, Simon and {Craig}, Matthew W. and {Cross}, Robert M. and {Cruz}, Kelle L. and {D'Eugenio}, Francesco and {Dencheva}, Nadia and {Devillepoix}, Hadrien A.~R. and {Dietrich}, J{\"o}rg P. and {Eigenbrot}, Arthur Davis and {Erben}, Thomas and {Ferreira}, Leonardo and {Foreman-Mackey}, Daniel and {Fox}, Ryan and {Freij}, Nabil and {Garg}, Suyog and {Geda}, Robel and {Glattly}, Lauren and {Gondhalekar}, Yash and {Gordon}, Karl D. and {Grant}, David and {Greenfield}, Perry and {Groener}, Austen M. and {Guest}, Steve and {Gurovich}, Sebastian and {Handberg}, Rasmus and {Hart}, Akeem and {Hatfield-Dodds}, Zac and {Homeier}, Derek and {Hosseinzadeh}, Griffin and {Jenness}, Tim and {Jones}, Craig K. and {Joseph}, Prajwel and {Kalmbach}, J. Bryce and {Karamehmetoglu}, Emir and {Ka{\l}uszy{\'n}ski}, Miko{\l}aj and {Kelley}, Michael S.~P. and {Kern}, Nicholas and {Kerzendorf}, Wolfgang E. and {Koch}, Eric W. and {Kulumani}, Shankar and {Lee}, Antony and {Ly}, Chun and {Ma}, Zhiyuan and {MacBride}, Conor and {Maljaars}, Jakob M. and {Muna}, Demitri and {Murphy}, N.~A. and {Norman}, Henrik and {O'Steen}, Richard and {Oman}, Kyle A. and {Pacifici}, Camilla and {Pascual}, Sergio and {Pascual-Granado}, J. and {Patil}, Rohit R. and {Perren}, Gabriel I. and {Pickering}, Timothy E. and {Rastogi}, Tanuj and {Roulston}, Benjamin R. and {Ryan}, Daniel F. and {Rykoff}, Eli S. and {Sabater}, Jose and {Sakurikar}, Parikshit and {Salgado}, Jes{\'u}s and {Sanghi}, Aniket and {Saunders}, Nicholas and {Savchenko}, Volodymyr and {Schwardt}, Ludwig and {Seifert-Eckert}, Michael and {Shih}, Albert Y. and {Jain}, Anany Shrey and {Shukla}, Gyanendra and {Sick}, Jonathan and {Simpson}, Chris and {Singanamalla}, Sudheesh and {Singer}, Leo P. and {Singhal}, Jaladh and {Sinha}, Manodeep and {Sip{\H{o}}cz}, Brigitta M. and {Spitler}, Lee R. and {Stansby}, David and {Streicher}, Ole and {{\v{S}}umak}, Jani and {Swinbank}, John D. and {Taranu}, Dan S. and {Tewary}, Nikita and {Tremblay}, Grant R. and {de Val-Borro}, Miguel and {Van Kooten}, Samuel J. and {Vasovi{\'c}}, Zlatan and {Verma}, Shresth and {de Miranda Cardoso}, Jos{\'e} Vin{\'\i}cius and {Williams}, Peter K.~G. and {Wilson}, Tom J. and {Winkel}, Benjamin and {Wood-Vasey}, W.~M. and {Xue}, Rui and {Yoachim}, Peter and {Zhang}, Chen and {Zonca}, Andrea and {Astropy Project Contributors}},
        title = "{The Astropy Project: Sustaining and Growing a Community-oriented Open-source Project and the Latest Major Release (v5.0) of the Core Package}",
      journal = {\apj},
         year = 2022,
        month = aug,
       volume = {935},
       number = {2},
          eid = {167},
        pages = {167},
          doi = {10.3847/1538-4357/ac7c74},
archivePrefix = {arXiv},
       eprint = {2206.14220},
 primaryClass = {astro-ph.IM},
       adsurl = {https://ui.adsabs.harvard.edu/abs/2022ApJ...935..167A}
}

@ARTICLE{2005MNRAS.361.1131F,
       author = {{Ferrario}, Lilia and {Wickramasinghe}, Dayal and {Liebert}, James and {Williams}, Kurtis A.},
        title = "{The open-cluster initial-final mass relationship and the high-mass tail of the white dwarf distribution}",
      journal = {\mnras},
         year = 2005,
        month = aug,
       volume = {361},
       number = {4},
        pages = {1131-1135},
          doi = {10.1111/j.1365-2966.2005.09244.x},
archivePrefix = {arXiv},
       eprint = {astro-ph/0506317},
 primaryClass = {astro-ph},
       adsurl = {https://ui.adsabs.harvard.edu/abs/2005MNRAS.361.1131F}
}

@ARTICLE{2014MNRAS.445.4287T,
       author = {{Tang}, Jing and {Bressan}, Alessandro and {Rosenfield}, Philip and {Slemer}, Alessandra and {Marigo}, Paola and {Girardi}, L{\'e}o and {Bianchi}, Luciana},
        title = "{New PARSEC evolutionary tracks of massive stars at low metallicity: testing canonical stellar evolution in nearby star-forming dwarf galaxies}",
      journal = {\mnras},
         year = 2014,
        month = dec,
       volume = {445},
       number = {4},
        pages = {4287-4305},
          doi = {10.1093/mnras/stu2029},
archivePrefix = {arXiv},
       eprint = {1410.1745},
 primaryClass = {astro-ph.SR},
       adsurl = {https://ui.adsabs.harvard.edu/abs/2014MNRAS.445.4287T}
}

@ARTICLE{2018MNRAS.476..496F,
       author = {{Fu}, Xiaoting and {Bressan}, Alessandro and {Marigo}, Paola and {Girardi}, L{\'e}o and {Montalb{\'a}n}, Josefina and {Chen}, Yang and {Nanni}, Ambra},
        title = "{New PARSEC data base of {\ensuremath{\alpha}}-enhanced stellar evolutionary tracks and isochrones - I. Calibration with 47 Tuc (NGC 104) and the improvement on RGB bump}",
      journal = {\mnras},
         year = 2018,
        month = may,
       volume = {476},
       number = {1},
        pages = {496-511},
          doi = {10.1093/mnras/sty235},
archivePrefix = {arXiv},
       eprint = {1801.07137},
 primaryClass = {astro-ph.SR},
       adsurl = {https://ui.adsabs.harvard.edu/abs/2018MNRAS.476..496F}
}

@ARTICLE{2026ApJ...996...69M,
       author = {{Miller}, David R. and {Caiazzo}, Ilaria and {Heyl}, Jeremy and {Richer}, Harvey B. and {Hollands}, Mark A. and {Tremblay}, Pier-Emmanuel and {El-Badry}, Kareem and {Rodriguez}, Antonio C. and {Vanderbosch}, Zachary P.},
        title = "{The White Dwarf Initial─Final Mass Relation from Open Clusters in Gaia DR3}",
      journal = {\apj},
         year = {2026a},
        month = jan,
       volume = {996},
       number = {1},
          eid = {69},
        pages = {69},
          doi = {10.3847/1538-4357/ae18c8},
archivePrefix = {arXiv},
       eprint = {2510.24877},
 primaryClass = {astro-ph.SR},
       adsurl = {https://ui.adsabs.harvard.edu/abs/2026ApJ...996...69M}
}

@ARTICLE{2006ApJ...643L.127W,
       author = {{Williams}, Kurtis A. and {Liebert}, James and {Bolte}, Michael and {Hanson}, Robert B.},
        title = "{A Hot DQ White Dwarf in the Open Star Cluster M35}",
      journal = {\apjl},
         year = 2006,
        month = jun,
       volume = {643},
       number = {2},
        pages = {L127-L130},
          doi = {10.1086/505211},
archivePrefix = {arXiv},
       eprint = {astro-ph/0604419},
 primaryClass = {astro-ph},
       adsurl = {https://ui.adsabs.harvard.edu/abs/2006ApJ...643L.127W}
}

@ARTICLE{2026ApJ..1007..183M,
       author = {{Miller}, David R. and {Heyl}, Jeremy and {Tremblay}, Pier-Emmanuel},
        title = "{Investigating the Observational Progenitor Mass Gap in the White Dwarf Initial-Final Mass Relation. I. Cluster Census and Characterization of the First White Dwarfs in the Gap}",
      journal = {\apj},
         year = {2026b},
        month = aug,
       volume = {1007},
       number = {2},
          eid = {183},
        pages = {183},
          doi = {10.3847/1538-4357/ae8b9e},
       adsurl = {https://ui.adsabs.harvard.edu/abs/2026ApJ..1007..183M}
}

@article{Miller2026Atypical,
   author = {{Miller}, David R. and {Caiazzo}, Ilaria and {Heyl}, Jeremy and {Tremblay}, Pier-Emmanuel},
   title = {Atypical White Dwarfs in Open Clusters},
   year = {2026c},
   journal = {arXiv e-prints},
   eid = {arXiv:2610.10699},
   eprint = {2610.10699},
   archivePrefix = {arXiv},
   primaryClass = {astro-ph.SR},
   doi = {10.48550/arXiv.2610.10699},
   note = {Submitted to ApJ}}

@ARTICLE{2016MNRAS.455.3413K,
       author = {{Kepler}, S.~O. and {Pelisoli}, I. and {Koester}, D. and {Ourique}, G. and {Romero}, A.~D. and {Reindl}, N. and {Kleinman}, S.~J. and {Eisenstein}, D.~J. and {Valois}, A.~D.~M. and {Amaral}, L.~A.},
        title = "{New white dwarf and subdwarf stars in the Sloan Digital Sky Survey Data Release 12}",
      journal = {\mnras},
         year = 2016,
        month = feb,
       volume = {455},
       number = {4},
        pages = {3413-3423},
          doi = {10.1093/mnras/stv2526},
archivePrefix = {arXiv},
       eprint = {1510.08409},
 primaryClass = {astro-ph.SR},
       adsurl = {https://ui.adsabs.harvard.edu/abs/2016MNRAS.455.3413K}
}

@ARTICLE{2023ApJ...956L..41M,
       author = {{Miller}, David R. and {Caiazzo}, Ilaria and {Heyl}, Jeremy and {Richer}, Harvey B. and {El-Badry}, Kareem and {Rodriguez}, Antonio C. and {Vanderbosch}, Zachary P. and {van Roestel}, Jan},
        title = "{An Extremely Massive White Dwarf Escaped from the Hyades Star Cluster}",
      journal = {\apjl},
         year = 2023,
        month = oct,
       volume = {956},
       number = {2},
          eid = {L41},
        pages = {L41},
          doi = {10.3847/2041-8213/acffc4},
archivePrefix = {arXiv},
       eprint = {2310.03204},
 primaryClass = {astro-ph.SR},
       adsurl = {https://ui.adsabs.harvard.edu/abs/2023ApJ...956L..41M}
}

@ARTICLE{2016arXiv161100036D,
       author = {{DESI Collaboration} and {Aghamousa}, Amir and {Aguilar}, Jessica and {Ahlen}, Steve and {Alam}, Shadab and {Allen}, Lori E. and {Allende Prieto}, Carlos and {Annis}, James and {Bailey}, Stephen and {Balland}, Christophe and {Ballester}, Otger and {Baltay}, Charles and {Beaufore}, Lucas and {Bebek}, Chris and {Beers}, Timothy C. and {Bell}, Eric F. and {Bernal}, Jos{\'e} Luis and {Besuner}, Robert and {Beutler}, Florian and {Blake}, Chris and {Bleuler}, Hannes and {Blomqvist}, Michael and {Blum}, Robert and {Bolton}, Adam S. and {Briceno}, Cesar and {Brooks}, David and {Brownstein}, Joel R. and {Buckley-Geer}, Elizabeth and {Burden}, Angela and {Burtin}, Etienne and {Busca}, Nicolas G. and {Cahn}, Robert N. and {Cai}, Yan-Chuan and {Cardiel-Sas}, Laia and {Carlberg}, Raymond G. and {Carton}, Pierre-Henri and {Casas}, Ricard and {Castander}, Francisco J. and {Cervantes-Cota}, Jorge L. and {Claybaugh}, Todd M. and {Close}, Madeline and {Coker}, Carl T. and {Cole}, Shaun and {Comparat}, Johan and {Cooper}, Andrew P. and {Cousinou}, M.-C. and {Crocce}, Martin and {Cuby}, Jean-Gabriel and {Cunningham}, Daniel P. and {Davis}, Tamara M. and {Dawson}, Kyle S. and {de la Macorra}, Axel and {De Vicente}, Juan and {Delubac}, Timoth{\'e}e and {Derwent}, Mark and {Dey}, Arjun and {Dhungana}, Govinda and {Ding}, Zhejie and {Doel}, Peter and {Duan}, Yutong T. and {Ealet}, Anne and {Edelstein}, Jerry and {Eftekharzadeh}, Sarah and {Eisenstein}, Daniel J. and {Elliott}, Ann and {Escoffier}, St{\'e}phanie and {Evatt}, Matthew and {Fagrelius}, Parker and {Fan}, Xiaohui and {Fanning}, Kevin and {Farahi}, Arya and {Farihi}, Jay and {Favole}, Ginevra and {Feng}, Yu and {Fernandez}, Enrique and {Findlay}, Joseph R. and {Finkbeiner}, Douglas P. and {Fitzpatrick}, Michael J. and {Flaugher}, Brenna and {Flender}, Samuel and {Font-Ribera}, Andreu and {Forero-Romero}, Jaime E. and {Fosalba}, Pablo and {Frenk}, Carlos S. and {Fumagalli}, Michele and {Gaensicke}, Boris T. and {Gallo}, Giuseppe and {Garcia-Bellido}, Juan and {Gaztanaga}, Enrique and {Pietro Gentile Fusillo}, Nicola and {Gerard}, Terry and {Gershkovich}, Irena and {Giannantonio}, Tommaso and {Gillet}, Denis and {Gonzalez-de-Rivera}, Guillermo and {Gonzalez-Perez}, Violeta and {Gott}, Shelby and {Graur}, Or and {Gutierrez}, Gaston and {Guy}, Julien and {Habib}, Salman and {Heetderks}, Henry and {Heetderks}, Ian and {Heitmann}, Katrin and {Hellwing}, Wojciech A. and {Herrera}, David A. and {Ho}, Shirley and {Holland}, Stephen and {Honscheid}, Klaus and {Huff}, Eric and {Hutchinson}, Timothy A. and {Huterer}, Dragan and {Hwang}, Ho Seong and {Illa Laguna}, Joseph Maria and {Ishikawa}, Yuzo and {Jacobs}, Dianna and {Jeffrey}, Niall and {Jelinsky}, Patrick and {Jennings}, Elise and {Jiang}, Linhua and {Jimenez}, Jorge and {Johnson}, Jennifer and {Joyce}, Richard and {Jullo}, Eric and {Juneau}, St{\'e}phanie and {Kama}, Sami and {Karcher}, Armin and {Karkar}, Sonia and {Kehoe}, Robert and {Kennamer}, Noble and {Kent}, Stephen and {Kilbinger}, Martin and {Kim}, Alex G. and {Kirkby}, David and {Kisner}, Theodore and {Kitanidis}, Ellie and {Kneib}, Jean-Paul and {Koposov}, Sergey and {Kovacs}, Eve and {Koyama}, Kazuya and {Kremin}, Anthony and {Kron}, Richard and {Kronig}, Luzius and {Kueter-Young}, Andrea and {Lacey}, Cedric G. and {Lafever}, Robin and {Lahav}, Ofer and {Lambert}, Andrew and {Lampton}, Michael and {Landriau}, Martin and {Lang}, Dustin and {Lauer}, Tod R. and {Le Goff}, Jean-Marc and {Le Guillou}, Laurent and {Le Van Suu}, Auguste and {Lee}, Jae Hyeon and {Lee}, Su-Jeong and {Leitner}, Daniela and {Lesser}, Michael and {Levi}, Michael E. and {L'Huillier}, Benjamin and {Li}, Baojiu and {Liang}, Ming and {Lin}, Huan and {Linder}, Eric and {Loebman}, Sarah R. and {Luki{\'c}}, Zarija and {Ma}, Jun and {MacCrann}, Niall and {Magneville}, Christophe and {Makarem}, Laleh and {Manera}, Marc and {Manser}, Christopher J. and {Marshall}, Robert and {Martini}, Paul and {Massey}, Richard and {Matheson}, Thomas and {McCauley}, Jeremy and {McDonald}, Patrick and {McGreer}, Ian D. and {Meisner}, Aaron and {Metcalfe}, Nigel and {Miller}, Timothy N. and {Miquel}, Ramon and {Moustakas}, John and {Myers}, Adam and {Naik}, Milind and {Newman}, Jeffrey A. and {Nichol}, Robert C. and {Nicola}, Andrina and {Nicolati da Costa}, Luiz and {Nie}, Jundan and {Niz}, Gustavo and {Norberg}, Peder and {Nord}, Brian and {Norman}, Dara and {Nugent}, Peter and {O'Brien}, Thomas and {Oh}, Minji and {Olsen}, Knut A.~G.},
        title = "{The DESI Experiment Part I: Science,Targeting, and Survey Design}",
      journal = {arXiv e-prints},
         year = 2016,
        month = oct,
          eid = {arXiv:1611.00036},
        pages = {arXiv:1611.00036},
          doi = {10.48550/arXiv.1611.00036},
archivePrefix = {arXiv},
       eprint = {1611.00036},
 primaryClass = {astro-ph.IM},
       adsurl = {https://ui.adsabs.harvard.edu/abs/2016arXiv161100036D}
}

@ARTICLE{2023MNRAS.526.4107P,
       author = {{Perren}, Gabriel I. and {Pera}, Mar{\'\i}a S. and {Navone}, Hugo D. and {V{\'a}zquez}, Rub{\'e}n A.},
        title = "{The Unified Cluster Catalogue: towards a comprehensive and homogeneous data base of stellar clusters}",
      journal = {\mnras},
         year = 2023,
        month = dec,
       volume = {526},
       number = {3},
        pages = {4107-4119},
          doi = {10.1093/mnras/stad2826},
archivePrefix = {arXiv},
       eprint = {2308.04546},
 primaryClass = {astro-ph.GA},
       adsurl = {https://ui.adsabs.harvard.edu/abs/2023MNRAS.526.4107P}
}

@ARTICLE{1992AJ....104.1876W,
       author = {{Weidemann}, Volker and {Jordan}, Stefan and {Iben}, Jr., Icko and {Casertano}, Stefano},
        title = "{White Dwarfs in the Halo of the Hyades Cluster: The Case of the Missing White Dwarfs}",
      journal = {\aj},
         year = 1992,
        month = nov,
       volume = {104},
        pages = {1876},
          doi = {10.1086/116364},
       adsurl = {https://ui.adsabs.harvard.edu/abs/1992AJ....104.1876W}
}

@ARTICLE{2026ApJ..1000..216K,
       author = {{Kilic}, Mukremin and {Bergeron}, Pierre and {Moss}, Adam and {Blouin}, Simon and {Green}, Matthew J. and {Jewett}, Gracyn and {Barrientos}, Manuel and {Albright}, Alexander L. and {Brown}, Warren R.},
        title = "{A Detailed Model Atmosphere Analysis of Hot White Dwarfs in DESI DR1}",
      journal = {\apj},
         year = 2026,
        month = apr,
       volume = {1000},
       number = {2},
          eid = {216},
        pages = {216},
          doi = {10.3847/1538-4357/ae43ee},
archivePrefix = {arXiv},
       eprint = {2602.07246},
 primaryClass = {astro-ph.SR},
       adsurl = {https://ui.adsabs.harvard.edu/abs/2026ApJ..1000..216K}
}

@ARTICLE{2010ApJ...720..581G,
       author = {{Gianninas}, A. and {Bergeron}, P. and {Dupuis}, J. and {Ruiz}, M.~T.},
        title = "{Spectroscopic Analysis of Hot, Hydrogen-rich White Dwarfs: The Presence of Metals and the Balmer-line Problem}",
      journal = {\apj},
         year = 2010,
        month = sep,
       volume = {720},
       number = {1},
        pages = {581-602},
          doi = {10.1088/0004-637X/720/1/581},
       adsurl = {https://ui.adsabs.harvard.edu/abs/2010ApJ...720..581G}
}

\end{document}